%% file: main.tex
\documentclass[acmsmall,screen]{acmart}
\AtBeginDocument{%
  }

\setcopyright{acmlicensed}
\copyrightyear{2025}
\acmYear{2025}
\acmDOI{XXXXXXX.XXXXXXX}
\acmConference[Conference acronym 'XX]{Make sure to enter the correct
  conference title from your rights confirmation email}{June 03--05,
  2025}{Woodstock, NY}

\acmISBN{978-1-4503-XXXX-X/2018/06}

\input{setup}

\begin{document}

\title{Every Maintenance Has Its Exemplar: The Future of Software Maintenance through Migration}

\author{Zirui Chen}
\orcid{0009-0004-6236-9150}
\affiliation{%
   \institution{The State Key Laboratory of Blockchain and Data Security, Zhejiang University}
   \city{Hangzhou}
  \country{China}
}
\email{chenzirui@zju.edu.cn}

\author{Xing Hu}
\orcid{0000-0003-0093-3292}
\affiliation{
   \institution{The State Key Laboratory of Blockchain and Data Security, Zhejiang University}
   \city{Hangzhou}
  \country{China}
}
\email{xinghu@zju.edu.cn}

\author{Xin Xia}
\orcid{0000-0002-6302-3256}
\authornote{Corresponding Author}
\affiliation{%
   \institution{The State Key Laboratory of Blockchain and Data Security, Zhejiang University}
   \city{Hangzhou}
  \country{China}
}
\email{xin.xia@acm.org}

\author{Xiaohu Yang}
\orcid{0000-0003-4111-4189}
\affiliation{%
   \institution{The State Key Laboratory of Blockchain and Data Security, Zhejiang University}
   \city{Hangzhou}
  \country{China}
}
\email{yangxh@zju.edu.cn}

\renewcommand{\shortauthors}{Chen et al.}

\begin{abstract}
    Maintenance is a critical stage in the software lifecycle, ensuring that post-release systems remain reliable, efficient, and adaptable. However, manual software maintenance is labor-intensive, time-consuming, and error-prone, which highlights the urgent need for automation. Learning from maintenance activities conducted on other software systems offers an effective way to improve efficiency. In particular, recent research has demonstrated that migration-based approaches transfer knowledge, artifacts, or solutions from one system to another and show strong potential in tasks such as API evolution adaptation, software testing, and migrating patches for fault correction. This makes migration-based maintenance a valuable research direction for advancing automated maintenance.
    
    This paper takes a step further by presenting the first systematic research agenda on migration-based approaches to software maintenance. We characterize the migration-based maintenance lifecycle through four key stages: \ding{182}  identifying a maintenance task that can be addressed through migration, \ding{183} selecting suitable migration sources for the target project,\ding{184} matching relevant data across systems and adapting the migrated data to the target context, and \ding{185} validating the correctness of the migration. We also analyze the challenges that may arise at each stage. Our goal is to encourage the community to explore migration-based approaches more thoroughly and to tackle the key challenges that must be solved to advance automated software maintenance.
\end{abstract}

\ccsdesc[500]{Software and its engineering~Maintaining software}
\ccsdesc[500]{Software and its engineering~Reusability}
\ccsdesc[300]{Software and its engineering~Automatic programming}

\keywords{Software maintenance, Software migration, Automated software engineering}


\maketitle

\input{tex/1-introduction}
\input{tex/2-related}
\section{Overview of Challenges}
\label{Challenge}

Software systems are becoming increasingly large, heterogeneous, and fast-evolving, which places higher demands on techniques capable of transferring maintenance knowledge across projects, ecosystems, and languages. In this section, we outline the key challenges for MBM research. \dx{For clarity, each identified challenge (denoted as \textit{``C''}) or opportunity (denoted as \textit{``O''}) is labeled with a unique index and a descriptive name (e.g., \textit{C1. Title of the challenge} and \textit{O1. Title of the opportunity}).}

\begin{figure}[htbp] 
  \centering	
  \includegraphics[width=0.98\linewidth]{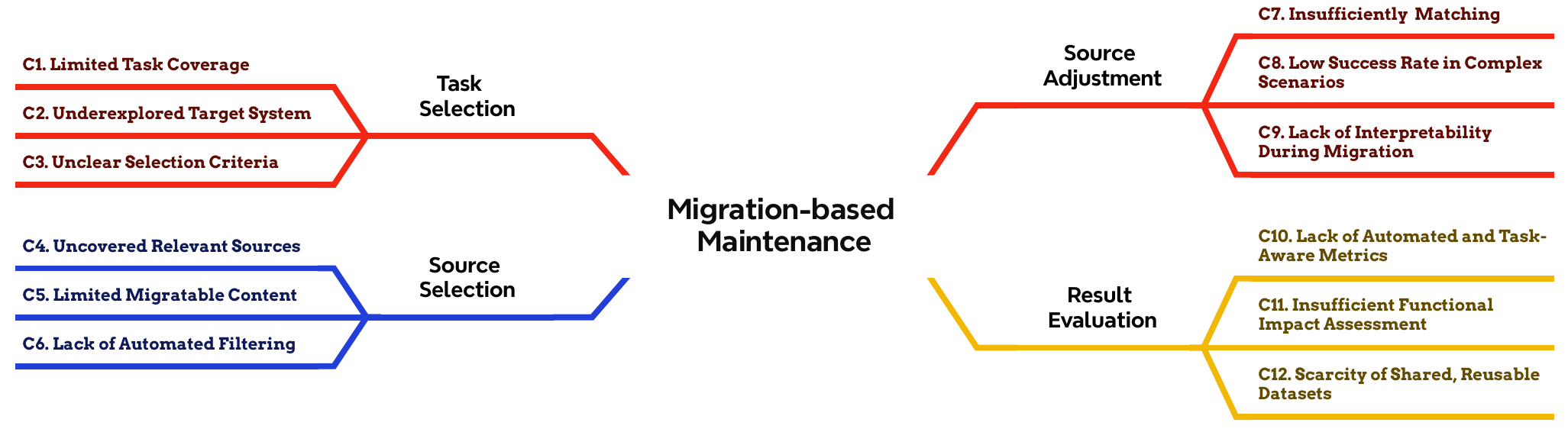}
  \caption{Overview of Challenges in MBM Process.}
  \label{fig: challenge}
  \Description{A mind map diagram titled Migration-based Maintenance illustrating challenges across four process stages.
  On the left, Task Selection branches into C1 (Limited Task Coverage), C2 (Underexplored Target System), and C3 (Unclear Selection Criteria);
  Source Selection branches into C4 (Uncovered Relevant Sources), C5 (Limited Migratable Content), and C6 (Lack of Automated Filtering).
  On the right, Source Adjustment branches into C7 (Insufficiently Matching), C8 (Low Success Rate in Complex Scenarios), and C9 (Lack of Interpretability During Migration);
  Result Evaluation branches into C10 (Lack of Automated and Task-Aware Metrics), C11 (Insufficient Functional Impact Assessment), and C12 (Scarcity of Shared, Reusable Datasets).}
\end{figure}

As illustrated in Figure~\ref{fig: challenge}, the challenges are organized along the four main stages of the MBM process: 
\textbf{Task Selection} (\textit{C1–C3}) captures the difficulty of choosing appropriate maintenance tasks for migration, such as limited coverage, underexplored target systems, and unclear selection criteria. 
\textbf{Source Selection} (\textit{C4–C6}) focuses on discovering relevant data sources, identifying migratable content, and filtering candidates. 
\textbf{Source Adjustment} (\textit{C7–C9}) highlights the complexity of accurately matching and adapting artifacts, as well as the lack of interpretability of migration operations. 
Finally, \textbf{Result Evaluation} (\textit{C10–C12}) deals with the challenges of automated and task-aware correctness metrics, assessing functional impact, and constructing reusable datasets for fair comparison. 
This structured view provides a roadmap for systematically addressing the obstacles that currently limit MBM research and practice. In the following sections, we present a detailed discussion of each challenge together with its corresponding research opportunities.

\input{tex/3-task}
\input{tex/4-source}
\input{tex/5-adaption}
\input{tex/6-evaluation}
\input{tex/7-perspective}

\section*{Acknowledgement}
This research is supported by the National Key R\&D Program of China (No.2024YFB4506400). We also thank the editors and anonymous reviewers for their insightful comments and suggestions.

\bibliographystyle{ACM-Reference-Format}
\bibliography{main}

\end{document}

%% file: setup.tex
\usepackage{makecell,tabularx}
\usepackage{enumitem}
\usepackage{pifont}
\usepackage{multirow}
\usepackage{multicol}

\newcommand{\dx}[1]{\textcolor{black}{{#1}}}

%% file: tex/1-introduction.tex
\section{Introduction}

Software maintenance is a cornerstone of the software lifecycle, encompassing activities such as correcting faults, improving performance, and modifying existing systems while preserving their integrity post-release~\cite{Bennett2000Maintenance}. Modern software, however, has become increasingly complex, with rapid release cycles, extensive third-party dependencies, support for multiple long-term supported (LTS) versions, and cloud-native architectures that involve hundreds of interdependent components. Developers must continually address API deprecations, respond to dependency upgrades, and integrate security patches~\cite{Fazzini2020Address}, all while minimizing service disruption. Maintenance already accounts for over half of the total lifecycle costs~\cite{Rajlich2000life}, and delays can lead to severe consequences, including missed business opportunities and security breaches~\cite{Bennett2000Maintenance}.

Manual maintenance under these conditions is labor-intensive, time-consuming, and error-prone~\cite{Lamothe2018Time, Linares2013Time}. This motivates the search for scalable, automated solutions capable of keeping pace with software maintenance. Although prior research has explored automation techniques, many are primarily data-driven. For example, automated program repair (APR) is often pre-trained on a large bug-fixing corpus and then fine-tuned to repair vulnerabilities ~\cite{Zhou2024Repair}, or guided by chains of thought (CoT) derived from previously successful fixes~\cite{Yin2025Repair}. These approaches rely on generalizing patterns learned from historical data to the current maintenance task. When such generalization fails, results can become unstable. For instance, Pan et al. ~\cite{Pan2024Patch} observed that large language models (LLMs) exhibit degraded performance on previously unseen maintenance tasks. These limitations highlight the absence of a general methodology that can be applied across maintenance tasks rather than designing a new solution or retraining a model for each task.

Recent research has explored transferring maintenance operations from one system to another to reduce the cost of automation~\cite{Gao2024Test, Zhang2024GUI, Mujahid2023Library}. Building on these efforts, we provide a definition of this line of work, which we term migration-based maintenance (MBM): the process of reusing concrete maintenance operations (e.g., patches, tests, configuration updates) that have already been applied in other systems and adapting them to the target project. \dx{Unlike learning-based approaches, which rely on historical data to predict potential solutions, MBM  ensures reliability by \ding{182} retrieving validated maintenance instances through a precise structural mapping process, and \ding{183} adapting these instances according to the syntactic and semantic constraints of the target system.} This reduces redundant engineering effort and avoids the generalization issues that often occur when the target task diverges from the training dataset. Existing studies have demonstrated the effectiveness of MBM in diverse contexts. For example, migrating tests from similar projects~\cite{Lin2022GUI, Zhang2024GUI, Gao2024Test}, reusing patches from fixed versions or forked projects to repair faults~\cite{Pan2024Patch, Ren2019Patch, Shi2023Patch, Nielsen2021Patch}, and leveraging solutions from projects using the same APIs to address API modifications~\cite{Lamothe2022API, Zhu2024API, Fazzini2019API}. These encouraging results motivate a deeper investigation into MBM.

In the coming years, MBM  is likely to play an increasingly important role in automated software maintenance. As software systems continue to grow larger, become more interconnected, and change at a faster pace, the demand for reusing and transferring maintenance knowledge will continue to grow. This calls for a clear research roadmap that goes beyond isolated, task-specific solutions and moves toward a common framework for reusing, adapting, and validating maintenance knowledge across projects.
In this paper, we outline a vision for future MBM research, pointing out the main challenges at each stage of the MBM process and highlighting research opportunities. Our goal is to encourage the development of MBM techniques that are interpretable, scalable, and easy to integrate into real-world development workflows, helping make automated software maintenance more practical and reliable.

Our paper is organized as follows:
Section \ref{Background} introduces the overall MBM process and clarifies the scope of our study.
Section \ref{Related} surveys existing research on MBM, summarizing representative approaches and key findings.
Section \ref{Challenge} provides an overview of the major challenges across all stages of MBM.
Sections \ref{Task}–\ref{Evaluate} discuss these challenges in detail and present potential opportunities for future work.
Finally, Section \ref{Perspective} offers a forward-looking perspective on MBM and highlights research directions.

%% file: tex/2-related.tex
\section{Background}
\label{Background}

In this section, we explain the concept of migration-based maintenance tasks and state the scope of our study. Moreover, we discuss the differences between MBM and other techniques.

\subsection{Stages of Migration-Based Maintenance}

\begin{figure}[htbp] 
  \centering	
  \includegraphics[width=12cm]{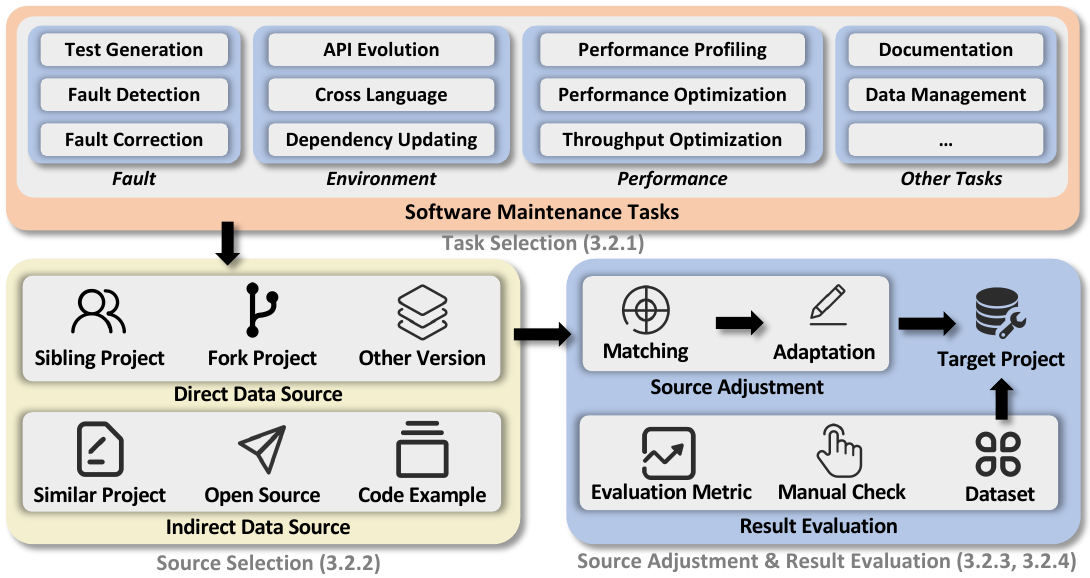}
  \caption{Overview of Migration-based Maintenance Process.}
  \Description{A diagram illustrating the four stages of the Migration-based Maintenance (MBM) process.
  The top section is 'Task Selection,' categorizing tasks into Fault (e.g., Test Generation, Fault Detection), Environment (e.g., API Evolution), Performance (e.g., Optimization), and Other Tasks.
  An arrow points down to 'Source Selection' on the left, which categorizes data sources into Direct (Sibling Project, Fork Project, Other Version) and Indirect (Similar Project, Open Source, Code Example).
  An arrow points right to the 'Source Adjustment' and 'Result Evaluation' sections. Source Adjustment shows a workflow from Matching to Adaptation, leading to the Target Project. Result Evaluation includes Evaluation Metrics, Manual Checks, and Datasets, which are used to validate the Target Project.}
  \label{fig: Migration}
\end{figure}

As illustrated in Figure~\ref{fig: Migration}, the MBM process can be systematically decomposed into four stages: task selection, source selection, source adjustment, and result evaluation. Together, these stages form a pipeline that guides how maintenance operation is identified, migrated, and validated across projects.

 Researchers need to identify concrete maintenance requirements during the \textbf{Task Selection} phase. 
Maintenance typically aims to correct faults, improve performance or other quality attributes, and adapt the software to evolving environments. 
Relevant tasks include supplementing tests for released software to catch latent faults, repairing confirmed defects, and responding to changes in dependencies, frameworks, and programming languages. Selecting which tasks to address is therefore not only the first step in the MBM process but also an open research problem. 
First, many potential maintenance tasks remain unexplored in the migration context. 
Second, maintenance tasks differ widely in difficulty.
Identifying which tasks are both high-value and technically feasible is crucial for guiding future MBM research.

 During the \textbf{Source Selection} phase, researchers need to choose appropriate migration data sources to support MBM tasks. The data sources are typically similar maintenance operations that have been manually conducted by other developers. Data sources for MBM tasks include a variety of existing maintenance artifacts. 
Already maintained versions of the same software often provide valuable information.
For instance, LTS releases of the Linux kernel contain security patches that can be migrated to other maintained branches. 
Fork or clone projects also carry relevant maintenance knowledge, including long-lived hard forks that have evolved independently or re-implementations in different languages of the same system. 
Functionally similar software can also reveal solutions to shared problems. 
Even code examples and snippets from open-source repositories or documentation can offer guidance for migration. Choosing a suitable source for migration is crucial in determining the efficacy of migration-based approaches.

\textbf{Source Adjustment} operations are required to align the selected data sources with the needs of the target project. 
This stage generally involves two steps. 
First, researchers perform a \emph{matching} process between the source and the target project. 
For example, in maintenance tasks related to API evolution, it is necessary to construct a mapping between the APIs used in the two projects.
For test migration, functional equivalence between the source and target tests must be established before the tests can be reused. 
Second, once matching is completed, the source artifacts must be \emph{modified}  to ensure that they function correctly in the target context. 
Direct migration is often not feasible, even between different versions of the same project, due to differences in APIs, configurations, or runtime environments. 
Existing research employs a variety of techniques to enable effective adaptation and ensure cross-project compatibility, such as heuristic rules and pre-trained models.

In the final \textbf{Result Evaluation} phase, researchers need to evaluate if the adapted migration operations have achieved the maintenance objectives, specifically assessing their impact on the current project functionalities. 
Errors, including compilation failures, may occur during the migration process, hindering the achievement of maintenance objectives. 
A comprehensive evaluation often requires measuring not only functional correctness but also side effects such as performance regressions, security implications, and maintainability of the resulting code. 
Different metrics have been proposed to validate the effectiveness of migration tasks for various scenarios, ranging from designing task-specific tests, run-time behavioral equivalence checking, to evaluating syntactic and semantic correctness. 
Despite these efforts, existing evaluation methods still depend heavily on manual effort and lack standardized benchmarks, making cross-study comparisons difficult.
Developing automated, interpretable, and reproducible evaluation frameworks remains an open research challenge and a key enabler for advancing MBM.

\subsection{Research Scope}

In this section, we delineate the scope of our work. 
Our study centers on software maintenance tasks addressed through migration-based approaches. 
To clearly situate our research focus, we further clarify how migration-based maintenance differs from development-phase activities and migration tasks during maintenance, 
and position it in relation to other techniques for automated software maintenance.

\subsubsection{Maintenance Tasks vs. Development/Evolution Tasks}

Some tasks, such as fault correction and software testing, occur in both software development/evolution and maintenance. 
Our study focuses exclusively on tasks performed during the maintenance phase, specifically tasks carried out for post-released software to correct faults, improve performance, and adapt to evolving environments.
In contrast, activities in other phases, such as automatic test generation for newly implemented features during the development phase, or early fault prediction and prevention techniques that aim to avoid defect introduction during evolution, are outside the scope of our study.

\subsubsection{Migration-Based Maintenance vs. Migration Maintenance Tasks}

In software maintenance, some tasks are explicitly defined as migration tasks, such as API migration~\cite{Collie2021API,Chen2020API} and test case migration~\cite{Gao2024Test}. 
Our focus is not on these task-specific migrations, but on migration as a methodological approach to enable automated maintenance, which refers to the reuse and adaptation of maintenance knowledge from other systems or versions to the target system. 
Unlike migration tasks that aim to move or transform specific artifacts, MBM generalizes this idea and applies it to a broader range of maintenance activities, including fault correction, API evolution response, and library updates.



\subsection{Comparison of MBM and Existing Techniques}

\dx{In this subsection, we compare MBM with established automated maintenance research, including search-based, constraint-based, template-based, and learning-based techniques. To provide a clear comparison, we select APR as a representative case study given its extensive presence in automatic maintenance research. In the following discussion, we contrast the core mechanisms, advantages, and limitations of each technique within the context of APR.}

\dx{As summarized in Table~\ref{tab:paradigm-comparison}, MBM distinguishes itself by directly leveraging validated maintenance operations from source projects. Since the solution logic in the reference source has already been confirmed to achieve the maintenance objective, MBM bypasses the computational overhead associated with extensive heuristic search or model training required by generation-based approaches. Furthermore, MBM exhibits strong generalizability, as it is not constrained by specific defect types or predefined templates. In terms of effectiveness, the adaptation of human-verified artifacts ensures high semantic fidelity and maintenance quality. However, the core challenge of MBM lies in its dependency on external resources, which means the feasibility is strictly related to the availability of a suitable reference source.}

\begin{table*}[t]
\centering
\caption{Comparison of MBM with Existing APR Techniques Identified in Recent Surveys~\cite{Huang2024APR}.}
\label{tab:paradigm-comparison}
\resizebox{\textwidth}{!}{%
\begin{tabular}{l|l|l|l|l}
\toprule
\textbf{Paradigm} & \textbf{Representative Works} & \textbf{Core Mechanism} & \textbf{Advantages} & \textbf{Limitations} \\ \midrule

\textbf{Search-based} & 
\begin{tabular}[c]{@{}l@{}}GenProg~\cite{LeGoues2012APR}, \\ SimFix~\cite{Jiang2018APR}\end{tabular} & 
\begin{tabular}[c]{@{}l@{}}\textbf{Heuristic Search}: Mutation \& \\ selection in a predefined space.\end{tabular} & 
\begin{tabular}[c]{@{}l@{}}Does not require external \\ training data.\end{tabular} & 
\begin{tabular}[c]{@{}l@{}}Search space explosion; \\ Invalid patches.\end{tabular} \\ \midrule

\textbf{Constraint-based} & 
\begin{tabular}[c]{@{}l@{}}Nopol~\cite{Xuan2017APR}, \\ SemFix~\cite{Nguyen2013APR}\end{tabular} & 
\begin{tabular}[c]{@{}l@{}}\textbf{Constraint Solving}: Synthesizing \\ code satisfying specifications.\end{tabular} & 
\begin{tabular}[c]{@{}l@{}}High repair quality for \\ specific bug types.\end{tabular} & 
\begin{tabular}[c]{@{}l@{}}Relies on strict specs; \\ Lack of flexibility.\end{tabular} \\ \midrule

\textbf{Template-based} & 
\begin{tabular}[c]{@{}l@{}}TBar~\cite{Liu2019APR}, \\ FixMiner~\cite{Koyuncu2020APR}\end{tabular} & 
\begin{tabular}[c]{@{}l@{}}\textbf{Pattern Matching}: Using pre- \\ defined/mined fix templates.\end{tabular} & 
\begin{tabular}[c]{@{}l@{}}Efficient; Reduces search \\ space significantly.\end{tabular} & 
\begin{tabular}[c]{@{}l@{}}Cannot fix bugs outside \\ defined patterns.\end{tabular} \\ \midrule

\textbf{Learning-based} & 
\begin{tabular}[c]{@{}l@{}}SequenceR~\cite{Chen2021APR}, \\ AlphaRepair~\cite{Xia2022APR}\end{tabular} & 
\begin{tabular}[c]{@{}l@{}}\textbf{Predictive Generation}: Learning \\ from defect repair samples data.\end{tabular} & 
\begin{tabular}[c]{@{}l@{}}Good generalization; \\ Handles complex logic.\end{tabular} & 
\begin{tabular}[c]{@{}l@{}}Data leakage/quality issues; \\ Hallucination risk.\end{tabular} \\ \midrule

\textbf{MBM} & 
\begin{tabular}[c]{@{}l@{}}PPatHF~\cite{Pan2024Patch}, \\ SKYPORT~\cite{Shi2023Patch}\end{tabular} & 
\begin{tabular}[c]{@{}l@{}}\textbf{Artifact Migration}: Adapting \\ validated artifacts from sources.\end{tabular} & 
\begin{tabular}[c]{@{}l@{}}\textbf{Verified Logic}: Preserves \\ verified logic from source.\end{tabular} & 
\begin{tabular}[c]{@{}l@{}}Dependent on availability of \\ homologous sources.\end{tabular} \\ \bottomrule
\end{tabular}%
}
\end{table*}

\section{State of the Research}
\label{Related}

In this section, we collect studies related to MBM and organize them by the four stages of the MBM process to outline the current research landscape.

\subsection{Literature Selection}

We define the following criteria to identify the most relevant literature regarding migration-based software maintenance. Our inclusion/exclusion criteria in the paper collection process include \textit{Screening} criteria and \textit{Relevance} criteria. As illustrated in Table \ref{tab: criteria}, the \textit{Screening} criteria are based on commonly used criteria for surveys and literature reviews ~\cite{Kitchenham2015Review} while the \textit{Relevance} criteria are based on the goal of our paper collection. With the criteria, we narrow our search results by examining each paper’s title, abstract, and content to obtain the most relevant papers.

Following Elder et al.~\cite{Elder2024Exploit}, we perform a keyword-based search across four major digital libraries: ACM Digital Library, IEEE Xplore, Compendex, and Web of Science. \dx{We construct our search strategy following the PICOC (Population, Intervention, Comparison, Outcomes, and Context) guidelines proposed by Kitchenham and Charters~\cite{PICOC} from the viewpoint of software engineering. Based on this framework, we formulate specific keywords for MBM. Table~\ref{tab:PICOC} summarizes the final search strings used in our paper collection.}

\dx{Guided by the objectives of software maintenance~\cite{ISO14764}, specifically \textbf{preventive actions} (e.g., test generation), \textbf{corrective actions} (e.g., patch generation, bug localization), \textbf{adaptive actions} (e.g., response to API evolution), and \textbf{perfective actions} (e.g.,  performance optimizing, functionality refining), we define four corresponding groups of keywords under the \emph{Outcome} and \emph{Context} dimensions of the PICOC framework and conduct searches for each group independently.} This process yields \dx{2,919} candidate papers. We then screen the titles and abstracts of these papers against our inclusion and exclusion criteria, resulting in a final set of \dx{38} papers relevant to MBM.

\begin{table}[t]
    \centering
    \caption{Inclusion and Exclusion Criteria.}
    \label{tab: criteria}
    \resizebox{0.9\linewidth}{!}{
    \begin{tabular}{l|l|l}
        \toprule
        & 
        \textbf{Inclusion Criteria:} Include papers if: & \textbf{Exclusion Criteria:} Exclude papers if:  \\ \midrule

        \rotatebox[origin=c]{90}{Screening}
         & 
       \multicolumn{1}{l|}{\begin{tabular}[c]{@{}p{6cm}@{}}
       
       - The paper must have an accessible full text.\\ 
       - The paper must be written in English. \\
       - The paper must have experimental results.\\
       - The paper is selected via peer review. \end{tabular}}

          &  
          \multicolumn{1}{l}{\begin{tabular}[c]{@{}p{6cm}@{}}
       - Studies belong to the following categories:  books, chapters, tutorials, technical reports. \\
       - The paper has less than four pages. \\
       - Duplicate papers.\end{tabular}} \\

        \midrule

        \rotatebox[origin=c]{90}{Relevance}
         & 
       \multicolumn{1}{l|}{\begin{tabular}[c]{@{}p{6cm}@{}}
       
       - The paper adopts migration to perform software maintenance. \\
       - The paper introduces previously nonexistent content to the software through migration.\end{tabular}}

          &  
          \multicolumn{1}{l}{\begin{tabular}[c]{@{}p{6cm}@{}}

        - The paper focuses on evaluating migrated code, rather than using migration as a means to achieve maintenance. \\
     \end{tabular}} \\

         \bottomrule
    \end{tabular}
    }
\end{table}

\begin{table}
    \caption{PICOC Criteria for Our Paper Collection.}
    \centering
    \resizebox{0.9\linewidth}{!}{
     \begin{tabular}{>{\centering\arraybackslash}p{2cm}|>{\centering\arraybackslash}p{3cm}|>{\centering\arraybackslash}p{8cm}}
      \toprule
       \textbf{  Category}    &  \textbf{ Main Term} & \textbf{  Search Terms} \\ \midrule
       \centering Population  & Software & ``software'' OR ``program'' OR ``code''\\ \midrule
       \centering Intervention  & Migration & ``migration'' OR ``porting''\\ \midrule
       \centering Comparison  & N.A. & N.A.\\ \midrule

     \multirow{5}{*}{Outcomes}

     & \multicolumn{1}{c|}{\dx{Preventive}} & 
     \multicolumn{1}{>{\centering\arraybackslash}p{8cm}}{\dx{``test'' OR ``validation''} }
     
     \\ [1pt]\cline{2-3}

      \rule{0pt}{10pt}
     &  \multicolumn{1}{c|}{\multirow{1}{*}{\dx{Corrective}}}  & \multicolumn{1}{>{\centering\arraybackslash}p{8cm}}{\dx{``patch'' OR ``detection'' OR ``fix'' OR ``repair''}} \\ [1pt]\cline{2-3}

     \rule{0pt}{10pt}
     & \multicolumn{1}{c|}{\dx{Adaptive}} & 
     \multicolumn{1}{>{\centering\arraybackslash}p{8cm}}{\dx{``adaption'' OR ``migration'' OR ``replacement'' OR ``upgrade''}} \\ [1pt]\cline{2-3} 

     \rule{0pt}{10pt}
     &  \multicolumn{1}{c|}{\multirow{1}{*}{\dx{Perfective}}}&  
     \multicolumn{1}{>{\centering\arraybackslash}p{8cm}}{ \dx{``improvement'' OR ``optimization'' OR ``enhancement''}} \\ 

        \midrule

        \multirow{5}{*}{Context}

     & \multicolumn{1}{c|}{\multirow{1}{*}{\dx{Preventive}}} & 
     \multicolumn{1}{>{\centering\arraybackslash}p{8cm}}{\dx{``test'' OR ``regression'' OR ``oracle''}} \\\cline{2-3}

     \rule{0pt}{10pt}
     
     & \multicolumn{1}{c|}{\multirow{1}{*}{\dx{Corrective}}}  & \multicolumn{1}{>{\centering\arraybackslash}p{8cm}}{\dx{``error'' OR ``bug'' OR ``fault'' OR ``defect'' OR ``vulnerability''}} \\ [1pt]\cline{2-3}

     \rule{0pt}{10pt}
     & \multicolumn{1}{c|}{\multirow{2}{*}{\dx{Adaptive}}} & 
     \multicolumn{1}{>{\centering\arraybackslash}p{8cm}}{\dx{``environment'' OR ``cross-platform'' OR ``API'' OR ``configuration'' OR ``portability''}} \\ [1pt]\cline{2-3} 

     \rule{0pt}{10pt}
     &  \multicolumn{1}{c|}{\multirow{1}{*}{\dx{Perfective}}}&  
     \multicolumn{1}{>{\centering\arraybackslash}p{8cm}}{ \dx{``performance'' OR ``feature''}}  \\ 

        \bottomrule
    \end{tabular}
    }
    \label{tab:PICOC}
\end{table}

\subsection{State of Existing Work}

\subsubsection{Task Selection}

\dx{As summarized in Table~\ref{tab:task}, existing studies have explored migration-based approaches across four major categories of maintenance tasks. 
First, for \textbf{preventive maintenance}, researchers have automated the migration of diverse test types, including unit tests~\cite{Gao2024Test, Mukelabai2023Test, Sondhi2022Test}, GUI tests~\cite{Zhang2024GUI, Talebipour2021GUI, Zhang2024GUI2}, and exploit tests~\cite{Jiang2023Exploit, Chen2024Exploit, Dai2021PoC, Zhou2024Exploit, zou2024syzbridge}. These automated migrations bring existing tests from external or prior codebases into the current project, allowing the capture of potential defects in advance and thereby supporting preventive maintenance. 
Regarding \textbf{corrective maintenance}, existing research leverages migration to accelerate both fault detection and repair. To assist bug localization, researchers use known defect locations from alternative language implementations~\cite{Tang2025mirror} or functionally similar systems~\cite{Guan2025DLbug} as guides to pinpoint bugs in the target project. Once located, defects are resolved by transplanting patches from historical versions~\cite{Shi2023Patch, Yang2023Patch}, or project forks~\cite{Ren2019Patch, Pan2024Patch}. In the case of Linux, migrating fixes from the mainline kernel to downstream distributions~\cite{Shariffdeen2021Patch} is also a form of preventive maintenance.
For \textbf{adaptive maintenance}, existing researches aim to resolve incompatibilities caused by library and framework updates through mining how other clients~\cite{Huang2024API, Ramos2023API, Nguyen2019API, Xu2019API} or usage examples~\cite{Lamothe2022API, Zhu2024API, Fazzini2019API} have adapted to the same API modifications, identifying the necessary changes, and automatically applying them to the target system. Finally, regarding \textbf{perfective actions}, MBM approaches enhance system maintainability by automating documentation generation. For instance, Dorninger et al.~\cite{Dorninger2017Documentation} implemented a re-documentation toolchain during a COBOL-to-Java migration, which automatically generates up-to-date documentation for the target system through migration.} 

\begin{table}[t]
  \centering
  \caption{Summary of Existing Migration Tasks by Data Source and Maintenance Purpose.}
  \resizebox{0.9\linewidth}{!}{%
  \begin{tabular}{cccccc}
    \toprule
    \textbf{Migration Data Source} & \textbf{\dx{Preventive Action}} & \textbf{\dx{Corrective Action}} & \textbf{\dx{Adaptive Action}} & \textbf{\dx{Perfective Action}}\\
    \midrule
    Cross Language Sibling & ~\cite{Talebipour2021GUI, Lin2022GUI, Qin2019Platform} & \dx{~\cite{Tang2025mirror}} & & \dx{~\cite{Dorninger2017Documentation}}\\
    Fork Project & ~\cite{Mukelabai2023Test, Mukelabai2021ForkTest} & \dx{~\cite{Pan2024Patch, Ren2019Patch, Bunyakiati2017Cherry}} & &\\
    Other Version &~\cite{Dai2021PoC, Saboor2025repair, Jiang2023Exploit} &~\cite{Shi2023Patch, Shariffdeen2021Patch, Yang2023Patch}  & \\
    
    Upstream Ecosystem &~\cite{Chen2024Exploit, Zhou2024Exploit, zou2024syzbridge} & & \dx{~\cite{2008migration}}  & \\
    
    Functionally Similar Project  & ~\cite{Gao2024Test, Sondhi2022Test, Zhang2024GUI, Zhang2024GUI2, Zhang2025chain, Vanover2020Test, Behrang2019GUI, Shen2025DL} & \dx{~\cite{Guan2025DLbug}} & \dx{~\cite{Xu2019API, Nguyen2019API, Mujahid2023Library, Alrubaye2019Library, Fazzini2020Address, Fazzini2019API}} & \\
   
    Code Example & & & \dx{~\cite{Almeida2024Oneshot, Lamothe2022API, Zhu2024API}} &  & \\
    \bottomrule
  \end{tabular}
  }
  \label{tab:task}
\end{table}

\subsubsection{Source Selection}

Data sources used in MBM can be broadly grouped by the strength of their relationship to the target system, which in turn affects the effort required for matching and adaptation.

\textbf{Strongly aligned sources.}
These sources have clear structural or version correspondences with the target, enabling direct reuse with minimal matching effort.
Examples include forked projects~\cite{Pan2024Patch}, upstream repositories~\cite{Zhou2024Exploit, zou2024syzbridge}, cross-language siblings~\cite{Talebipour2021GUI, Lin2022GUI}, and previously maintained versions~\cite{Dai2021PoC, Saboor2025repair, Jiang2023Exploit}.
Such relationships provide natural anchors (e.g., shared identifiers, similar file layouts), making tasks such as patch porting, exploit migration, or test transfer comparatively straightforward.

\textbf{Loosely aligned sources.}
These sources exhibit weaker structural similarity and require additional discovery or semantic matching before reuse.
Typical cases include functionally similar projects~\cite{Gao2024Test, Sondhi2022Test, Zhang2024GUI, Alrubaye2019Library, Mujahid2023Library},
where researchers must build cross-project mappings, and evolving code examples~\cite{Zhu2024API, Lamothe2022API},
where the relevant usage snippets need to be retrieved, aligned, and adapted to the target context.

\subsubsection{Source Adjustment}

This stage comprises two  activities: 
\ding{182} Matching entities between the source and target systems and 
\ding{183} Adapting the matched artifacts to ensure their correct functionality after migration.

For sources with high structural similarity to target projects, such as forked projects and previously maintained versions, allow relatively direct matching. 
For instance, Pan et al.~\cite{Pan2024Patch} manually identified corresponding functions for patch porting across forks, 
while Dai et al.~\cite{Dai2021PoC} leveraged method names and structural similarity to detect refactorings across versions. 
Upstream–downstream projects further benefit from explicit dependency relationships that narrow the search space for potential matches~\cite{Chen2024Exploit, Zhou2024Exploit}. In contrast, sources with significant structural divergence, such as cross-language siblings and functionally similar projects, 
require more sophisticated matching strategies to construct reliable correspondences. 
Gao et al.~\cite{Gao2024Test} combined token- and semantic-level similarity to align classes and functions across Java and C++ projects, 
whereas Huang et al.~\cite{Huang2024API} employed semantic embeddings to reveal potential API correspondences. 
In the context of GUI test migration, Qin et al.~\cite{Talebipour2021GUI} computed cosine similarity over widget labels to match events between iOS and Android applications, 
and Lin et al.~\cite{Lin2022GUI} encoded GUI screens to support cross-platform mapping.

To ensure the executability of migrated content, researchers employ various adjustment techniques. Random-based techniques are widely used for adjusting the source. Researchers extract path or state information from the data source and employ strategies like greedy-random~\cite{Zhang2024GUI}, fuzz testing~\cite{Zhou2024Exploit, Elder2024Exploit, Jiang2023Exploit}, and genetic algorithm tools~\cite{Chen2024Exploit} to replicate this state on the target project for successful migration. Adjusting the data source based on rules has also been employed by previous researchers~\cite{Gao2024Test, Chen2024Exploit}. Recently, significant potential has been shown by LLMs in migration tasks, where researchers collect instances of the target task and fine-tune LLMs to accomplish uncommon migration tasks~\cite{Pan2024Patch}.

\subsubsection{Result Evaluation}

\begin{table}[t]
\centering
\caption{Evaluation metrics and datasets used in MBM studies.}
\label{tab:api-updating-eval}
\resizebox{\linewidth}{!}{%
\begin{tabular}{c|c|c|l}
\toprule
\textbf{Object} & \textbf{Approach} & \textbf{Dataset} & \textbf{Evaluation Metric} \\
  \midrule
\multirow{5}{*}{\shortstack{\dx{Preventive}}} 
 & TestMig~\cite{Talebipour2021GUI} & 5 popular mobile applications (iOS \& Android) & Successful Execution Rate of Migrated Tests \\
 & MUT~\cite{Gao2024Test} & 5 prevalent functional domain projects & Success Rate (Manual Validation) \\
 & AEM~\cite{Jiang2023Exploit} & 67 exploit migration cases & Migration Success Rate (Manual Validation) \\
 & METALLICUS~\cite{Sondhi2022Test} & 67 defects from 42 project pairs & Precision \& Recall of Generated Tests \\
 & Mukelabai et al.~\cite{Mukelabai2023Test} & 305 fork ecosystems on GitHub & Test Reuse Rate (Manual Validation) \\
 \midrule

\multirow{5}{*}{\shortstack{\dx{Corrective}}} 
& Guan et al.~\cite{Guan2025DLbug} & 638 bugs from four projects (Java \& C\#) & - \\
 & CrossProbe~\cite{Tang2025mirror} & 182 issues from two DL libraries &  Bug Localization (Manual Inspection) \\
 & PPatHF~\cite{Pan2024Patch} & 310 Neovim patches ported from Vim & Edit Distance (Human-ported Patch) \\
 & \dx{FixMorph~\cite{Shariffdeen2021Patch}} & \dx{350 Linux mainline patches} & \dx{Syntactic \& Semantic Equivalence}\\
 & \dx{TSBPORT~\cite{Yang2023Patch}} & \dx{1,815 real-world Linux security patch} & \dx{Success Rate (Manual Validation)} \\

  \midrule
\multirow{3}{*}{\shortstack{\dx{Adaptive}}} 
 & Meditor~\cite{Xu2019API} & 4 Java libraries, 751 commits & Accuracy (Manual Inspection) \\
 & FOCUS~\cite{Nguyen2019API} & 5,147 randomly selected Java projects & Top-N Accuracy \\
 & Alrubaye et al.~\cite{Alrubaye2019Library} & 57,447 OSS Java projects, 9 library migrations & Accuracy (Manual Inspection) \\

   \midrule
\multirow{1}{*}{\shortstack{\dx{Perfective}}} 
 & \dx{Dorninger et al.}~\cite{Dorninger2017Documentation} & \dx{-} & \dx{-} \\

\bottomrule
\end{tabular}%
}
\end{table}

\dx{To assess the prevailing evaluation metrics in MBM, for each task and source, we highlight representative studies to illustrate common dataset choices and evaluation strategies. Table~\ref{tab:api-updating-eval} provides an overview of how different MBM tasks adopt strategies for dataset construction and metric design.}

For dataset construction, API updating research typically collects commit histories from a moderate number of libraries or projects and manually annotates API mappings to build ground truth~\cite{Xu2019API, Nguyen2019API, Alrubaye2019Library}.
Bug localization studies curate labeled issues and bug reports from real-world projects, often focusing on cross-language or multi-project cases to test generalizability~\cite{Guan2025DLbug, Tang2025mirror}.
Patch generation research frequently mines manually backported or cherry-picked commits from ecosystems such as Linux or Vim–Neovim, providing a natural oracle for correct patch content~\cite{Shariffdeen2021Patch, Pan2024Patch, Yang2023Patch}.
Test generation and migration studies typically source tests from fork ecosystems~\cite{Mukelabai2023Test}, GUI event sequences from mobile apps~\cite{Talebipour2021GUI}, or functional domain projects~\cite{Gao2024Test}, enabling evaluation of coverage and executability.

For metrics, current studies largely rely on task-specific, often human-intensive measures.
API updating commonly reports Top-$k$ accuracy or Mean Reciprocal Rank (MRR), reflecting the ranking quality of recommended API mappings~\cite{Huang2024API, Lamothe2022API}.
Patch migration studies compare generated patches against developer-ported ones, measuring syntactic and semantic similarity or manual success rates~\cite{Pan2024Patch, Shariffdeen2021Patch, Yang2023Patch}.
Bug localization results are judged by whether the localized statements or files overlap with ground-truth fixes.
Test migration relies heavily on manual inspection or execution success rates to verify correctness~\cite{Zhang2024GUI, Talebipour2021GUI}, and sometimes measures productivity gains by counting steps or time saved relative to manual test authoring~\cite{Gao2024Test}.
Exploit migration tasks use successful triggering of vulnerabilities as the evaluation signal~\cite{Chen2024Exploit, zou2024syzbridge}.

Overall, while these datasets and metrics enable targeted evaluation, they also reveal two gaps: \ding{182} many tasks still depend on costly manual validation, limiting reproducibility and scalability, and \ding{183} metrics often emphasize syntactic similarity or ranking quality rather than confirming that the maintenance goal has been truly achieved.

%% file: tex/3-task.tex
\section{Task Selection}
\label{Task}

Task selection for MBM faces three major challenges: \textit{Limited Task Coverage}, \textit{Underexplored Target Systems}, and \textit{Unclear Selection Criteria}.
Current work has mostly focused on a narrow range of tasks such as patch porting, test migration, and API updating, leaving performance optimization, configuration evolution, and documentation maintenance largely untouched.
Moreover, most studies concentrate on traditional software systems (e.g., mobile apps, web services, OS kernels), while emerging domains such as deep learning systems, protocols, and smart contracts remain underexplored despite their pressing maintenance needs.
Finally, the field lacks systematic criteria for deciding when migration is feasible, which leads to wasted effort when candidate sources are semantically incompatible with the target task.

\subsection{C1. Limited Task Coverage}
Current MBM research has concentrated on a narrow subset of maintenance tasks, primarily fault correction through patch porting, test migration, and API evolution handling. However, the software maintenance landscape is considerably broader. Tasks such as performance optimization, configuration evolution, and documentation maintenance remain largely unexplored from a migration perspective.  For instance, many systems undergo repeated performance improvement through code refactoring or concurrency adjustments, yet there are no established methods to migrate these optimizations across projects or versions.

\dx{As illustrated in Table \ref{tab:task_taxonomy}, to provide a systematic understanding of the task coverage of MBM, we categorize its applicable maintenance tasks into three categories: \ding{182} Logic-Shared Tasks, \ding{183} Environment-Driven Tasks, and \ding{184} Consistency-Preserving Tasks.}

\begin{table*}[t]
\centering
\caption{Categorization of MBM Applicable Tasks.}
\label{tab:task_taxonomy}
\small 

\begin{tabularx}{\linewidth}{@{} 
    >{\raggedright\arraybackslash}p{2.3cm} 
    >{\raggedright\arraybackslash}p{3.8cm} 
    X @{}}
\toprule
\textbf{Category} & \textbf{Definition Summary} & \textbf{Typical Maintenance Tasks} \\
\midrule

\textbf{Logic-Shared Tasks} & 
Maintenance driven by changes in source projects, which is transferable to target projects via a shared solution. & 
\begin{itemize}[leftmargin=*, nosep, after=\vspace{-\baselineskip}, before=\vspace{-0.5\baselineskip}]
    \item \textbf{Test Case Augmentation}: Transplanting tests from functionally similar projects or cross-language twins.
    \item \textbf{Similar Bug Fixing}: Applying fix patterns found in other projects implementing identical algorithms.
\end{itemize} \\ 
\midrule

\textbf{Environment-Driven Tasks} & 
Maintenance necessitated by external environmental changes, which is resolved by adapting patterns from source projects sharing the same environment. & 
\begin{itemize}[leftmargin=*, nosep, after=\vspace{-\baselineskip}, before=\vspace{-0.5\baselineskip}]
   \item \textbf{CI/CD Configuration Repair}: Updating scripts (e.g., GitHub Actions) due to platform changes by referencing source projects that have successfully adapted.
    
    \item \textbf{API Adaptation}: Modifying API usages to resolve library/framework upgrades according to updated usages in migrated projects.
\end{itemize} \\  
\midrule

\textbf{Consistency-Preserving Tasks} & 
Maintenance propagated across different versions, forks, or variants of the same project lineage to ensure consistency. & 
\begin{itemize}[leftmargin=*, nosep, before=\vspace{-0.5\baselineskip}]
       \item \textbf{Backporting/Synchronizing Patches}: Applying security fixes from main branches to release branches or between forks. 
    
        \item \textbf{Test Suite Synchronization}: Propagating test cases from latest releases to maintenance versions to ensure behavioral consistency.
        
\end{itemize} \\  

\bottomrule
\end{tabularx}
\end{table*}

\begin{itemize}

\item {\textbf{Logic-Shared Tasks.} \dx{This category encompasses maintenance tasks driven by the similarities between the source and target projects. It is common for different projects to implement identical functionalities (e.g., utility modules for URL parsing, JSON serialization ). MBM is particularly suitable for this category because the shared logic implies a shared solution space. For example, when a test case augmentation is validated in a source project, it serves as a proven reference for functionally similar targets. The primary advantage here is the reduction of uncertainty by adapting existing solutions that have already withstood real-world usage. We provide an example for this category in O1}.}

\item {\textbf{Environment-Driven Tasks.} \dx{This category addresses maintenance tasks necessitated by changes in the external environment rather than internal business requirements. Software relies heavily on external environments, such as third-party APIs, CI/CD platforms, and system configurations. Consequently, the evolution of these environments drives the need for adaptive maintenance to ensure continued compatibility. MBM is particularly suitable for such tasks, as it leverages existing adaptation instances from the community to provide a timely and practical solution. This is detailed in O2, which showcases the superiority of MBM in resolving cases that are overlooked by rule-based approaches.}}

\item{ \textbf{Consistency-Preserving Tasks.} \dx{This category targets maintenance activities across different versions, forks, or variants of the same project. Unlike cross-project migration, the source and target here share a common ancestry. Consequently, maintenance efforts invested in a specific source can be frequently reused in other variants due to structural similarities, necessitating propagated maintenance. MBM is inherently effective in this case because a change applied to one version acts as a strong signal for other active versions. We discuss this in O3, showing how MBM can automatically solve outdated documentation with code evolution.}}

\end{itemize}

\textbf{O1. Performance Optimization.} How can we systematically migrate performance improvements across versions or related projects to reduce repeated tuning effort?
Performance optimization is a frequent and costly maintenance activity, involving refactoring~\cite{Traini2021refactor}, concurrency adjustments~\cite{Lin2014Refactor}, and compiler-level tweaks~\cite{Khatchadourian2024Efficiency}.
Currently, such optimizations are often applied in isolation, meaning that similar systems or sibling projects repeatedly undergo the same performance tuning process from scratch.
Migration-based approaches hold the promise of automatically capturing these optimizations, e.g., through mining performance-improving commits, profiling results, or configuration diffs, and reapplying them in similar contexts.
The key insights include automatically distinguishing functional fixes from performance-oriented changes, generalizing optimization patterns across different codebases or environments, and validating that migrated optimizations indeed improve performance under the new workload.
We envision approaches that combine program-difference analysis, performance profiling, and learned models to transfer optimizations with minimal manual intervention, ultimately enabling performance-aware maintenance at scale.

For example, recent work on KernelBench~\cite{KernelBench} explores the use of LLMs to automatically generate optimized GPU kernels for PyTorch workloads, achieving both correctness and significant speedups for some operators. 
For example, in PyTorch, the Gaussian Error Linear Unit (GELU) is implemented as:
\[
\text{GELU}_{\text{Torch}}(x) = 0.5 \cdot x \cdot \bigl(1 + \tanh\bigl(\sqrt{2/\pi}(x + 0.044715x^{3})\bigr)\bigr),
\]
and KernelBench-generated kernels optimize this computation by fusing operations into a single CUDA kernel and applying constant folding (e.g., precomputing $\sqrt{2/\pi}=0.7978845608028654f$ to avoid redundant runtime computation).
In TensorFlow, GELU is defined as:
\[
\text{GELU}_{\text{TF}}(x) = 0.5 \cdot x \cdot \bigl(1 + \mathrm{erf}(x / \sqrt{2})\bigr).
\]
While mathematically equivalent, the implementations differ, leaving room for learned optimizations (such as operator fusion and constant folding) discovered in PyTorch to be systematically migrated to TensorFlow or other frameworks.
For example, \texttt{tf.math.erf} repeatedly computes $\sqrt{2}$ at runtime. 
A concrete migration opportunity is to port the optimization strategy learned from PyTorch to TensorFlow by folding $\sqrt{2}$ into a precomputed constant and generating a fused kernel that combines the elementwise operations. 
This cross-framework migration could reduce redundant tuning efforts and help achieve performance parity across ML stacks.

\dx{Compared to data-driven approaches, the distinct advantage of MBM lies in its ability to explicitly determine where performance improvements are needed and how they should be applied. Data-driven methods operate probabilistically as they lack certainty regarding whether a specific location is maintainable, carrying the risk of missing valid maintenance opportunities. MBM mitigates this uncertainty by leveraging solutions where the maintenance location and outcome are pre-validated in the source (GELU in PyTorch), which allows MBM to directly pinpoint the maintenance location and synthesize a strategy with proven logic (GELU in TensorFlow).}

\textbf{O2. Configuration Update.}  
Modern software systems heavily rely on configuration files, environment variables, and deployment descriptors to control their runtime behavior. These configurations often evolve alongside the system to adapt to new environments, security requirements, and performance goals. However, manual propagation of configuration updates across different versions, environments, or forks is error-prone and time-consuming, frequently leading to misconfigurations that cause outages or vulnerabilities. Migration-based approaches offer a promising solution: by extracting configuration changes from a maintained source system and systematically applying them to other systems, we can reduce configuration drift and improve reliability. Future research should focus on learning robust configuration mappings, handling semantic differences between environments, and automatically validating migrated configurations through runtime or static analysis.

Existing research explores configuration update rules through rule-based approaches. For example, Hossain et al.~\cite{CIgrate} investigated the problem of semantically preserving CI configuration migration across different platforms. They proposed transformation rules to convert YAML-based configurations between CI services while retaining the intended behavior. However, they noted that some Travis CI features (e.g., \texttt{addons} or \texttt{sudo}) lack direct equivalents in GitHub Actions, which leads to missing functionality or necessitates manual adjustments after migration. \dx{Since no predefined rule existed for these edge cases, their method resulted in broken workflows or required manual intervention.}

\dx{In contrast, MBM overcomes this limitation by leveraging specific, validated exemplars rather than relying on abstract rules. Although a general mapping for sudo may not exist, individual open-source projects have successfully devised solutions. For instance, the HTTP Toolkit project~\cite{Toolkit} successfully migrated its complex sudo usage by re-architecting its jobs into equivalent GitHub workflows. Similarly, Flipt~\cite{OpenSourceGo} completed a verified migration of their pipeline. For MBM, researchers should design algorithms that automatically identify and match these suitable migration exemplars. By treating these successful migration instances as reference sources, MBM can transfer existing configuration-migration logic to the target project, thereby addressing complex cases that fall outside general transformation rules.}

\textbf{O3. Documentation Evolution.} Automatically propagating documentation updates across versions, forks, and related projects is essential to maintain consistency and accuracy. 
Software documentation evolves alongside code, but in practice it often lags behind~\cite{Tang2025documentation}, leading to inconsistencies that confuse developers and users.
When software is refactored or semantics change, corresponding updates to documentation are frequently manual, error-prone, and easy to overlook.
Migration-based approaches could systematically detect documentation change and suggest or directly apply equivalent changes in sibling projects, forks, or other versions.
The main insight lies in aligning documentation entities with their code counterparts, generalizing updates across different documentation styles and languages, and verifying the correctness of migrated text.
Recent advances in NLP and LLM-based summarization offer promising opportunities to automate this migration process.
Such techniques could not only reduce stale or misleading documentation but also improve developer productivity by keeping guidance synchronized with the system after maintenance.

For example, an issue for Python~\cite{pythonIssue} reported that the documentation for the \texttt{`-3'} command-line option in Python~2.7 was outdated. 
It listed a set of incompatibilities handled by \texttt{lib2to3} but failed to include the \texttt{reload} fixer introduced in Python~3.4, causing confusion for developers preparing migration. This highlights the need for automated approaches to keep documentation synchronized with code evolution. 
MBM techniques could address this by mining documentation changes in newer versions
and automatically suggesting or propagating equivalent updates to older branches, ensuring consistency across versions.

\subsection{C2. Underexplored Target Systems}
Most existing MBM studies have focused on traditional software systems such as mobile applications, web services, and operating system kernels. 
However, the software landscape has expanded to include other systems where maintenance needs are just as critical but substantially challenging. 
Examples include deep learning systems, protocols, and smart contracts. 
Deep learning systems require co-evolution of data and training code, making regression testing and cross-framework consistency checks challenging. 
Protocols must remain backward compatible with deployed clients while incorporating bug fixes or security enhancements, posing a significant challenge during maintenance. 
Smart contracts pose a unique challenge for maintenance because they are stateful programs with complex interdependent functionalities, making it difficult and expensive to manually create test cases that cover realistic usage scenarios. 
Exploring MBM in these emerging systems holds potential for addressing domain-specific challenges that remain largely unstudied.

\textbf{O4. Migration on Smart Contract.}  \dx{The applicability of MBM on smart contracts is directly enabled by the Data Transparency and Widespread Code Reuse of the blockchain ecosystem. First, unlike traditional software, where runtime logs are private, the public availability of execution history on the blockchain provides a rich data source for MBM. MBM approaches can migrate and replay these real-world transactions from a deployed contract to a candidate upgrade for supporting regression testing. Second, the widespread code reuse in blockchain ecosystems also enables MBM. For example, a vast number of smart contracts rely on identical standard libraries or token implementations to ensure interoperability and security. MBM leverages this characteristic to transfer maintenance operations such as security patches and gas optimizations from a primary protocol to its derivatives.}

\textbf{O5. Migration on Protocols.}  Protocols define the communication rules and data formats that enable distributed systems to interoperate. Maintaining protocol implementations is challenging because upgrades must remain backward compatible with existing clients and services, while addressing security flaws or extending functionality. Incompatibilities may result in service outages or consensus failures, especially in decentralized or large-scale systems. Migration-based approaches could significantly improve protocol maintenance. For example, network traces or message logs collected under an old protocol version can be transformed into regression tests and replayed against upgraded implementations to validate interoperability. Similarly, bug-fixing patches for one implementation of a protocol could be adapted and ported to other independent implementations to ensure ecosystem-wide consistency and security. Future research could explore automated log-to-test generation, semantic diffing of protocol specifications, and cross-implementation patch migration techniques to maintain reliability in evolving protocol ecosystems.

\textbf{O6. Migration on the DL Ecosystem.}  The deep learning ecosystem consists of frameworks such as PyTorch, TensorFlow, and JAX, as well as DL compilers like TVM and TensorRT, together with heterogeneous backends such as GPU, TPU, and custom accelerators. These components form a complex software stack where reliability and performance are critical. Recent research has already explored test migration between DL frameworks~\cite{Guan2025DLbug} and test migration from frameworks to compiler frontends~\cite{Shen2025DL}, revealing inconsistencies across different components of the DL ecosystem.
However, a wide range of migration opportunities remains underexplored.
For instance, test migration between compiler frontends could uncover inconsistencies.
Patch and optimization migration between frameworks could accelerate bug fixing and share tuning knowledge, such as quantization or scheduling strategies.
Cross-backend test migration could systematically evaluate functional and performance consistency across GPUs, TPUs, and other accelerators. Exploring these directions would enable MBM to systematically reduce duplicated debugging and optimization effort, improve cross-stack reliability, and enhance the maintainability of the DL software ecosystem.

\textbf{O7. Migration on Framework Evolutions.}
Framework replacement is a common yet challenging maintenance scenario. When a core framework reaches end-of-life or becomes obsolete, projects must transition to a new, actively maintained framework to ensure long-term sustainability. Real-world examples include web front-end migrations such as AngularJS (deprecated since 2022) to React or Vue, or backend migrations like Struts to Spring MVC. Such transitions are costly because they require translating framework-specific APIs, configuration files, and idioms into their counterparts in the new framework while preserving system behavior. Migration-based approaches offer a promising solution by learning from existing projects that have already completed similar migrations. For instance, API-level mappings, component usage patterns, and configuration transformations can be mined from these projects and generalized into reusable migration rules. These learned rules can then be applied to automate portions of the framework migration process, reducing manual rewriting effort.

\subsection{C3. Unclear Selection Criteria}
Current MBM research typically assumes that the target maintenance task is suitable for migration-based approaches, focusing primarily on how to match and adapt source artifacts once a feasible source is available. However, in practice, determining whether a given task pair is suitable for MBM is a challenging problem.
Reference data may be unavailable, outdated, or semantically distant from the target project, leading to wasted effort if the migration fails.
Moreover, even when candidate sources exist, it remains unclear how to evaluate their relevance and the cost–benefit trade-off of migration compared to manual implementation.
The absence of systematic selection criteria hinders the practical adoption of MBM in real-world maintenance workflows.

\textbf{O8. Migration Feasibility Prediction.}
A persistent gap in MBM practice is the lack of predictive mechanisms to estimate migration success before execution.
Many current studies implicitly assume the availability of reusable sources or similar tasks, which often do not hold in practice. This leads to wasted computational and human resources when migration proves infeasible due to fundamental incompatibilities, such as missing dependencies, semantic divergences, or architectural drift between projects.
Migration feasibility prediction aims to address this by automatically estimating the likelihood that a maintenance task can be successfully migrated.
Promising research directions include learning predictive models from historical migration outcomes, quantifying structural and semantic similarity between source and target systems, and developing early-warning indicators for migration risks (e.g., predicting the likelihood of compilation failure or behavioral divergence).
Accurate feasibility prediction could guide developers toward cost-effective maintenance strategies and help allocate resources to tasks with a higher probability of successful automation.

\textbf{O9. Improving Existing Migration Techniques.}
Beyond determining which tasks are suitable for MBM, there is a strong opportunity to systematically improve the success rate and quality of migration techniques.
Existing approaches often rely on heuristic rules or partial automation~\cite{Gao2024Test}, which can result in incomplete migrations, compilation errors, or behavioral inconsistencies.
The emergence of LLMs offers new opportunities to enhance both the automation and the accuracy of MBM, showing promise in identifying transferable artifacts, performing source–target matching, and executing the migration itself.
For example, LLMs could infer candidate migrations, which are then validated through type checking, compilation, or differential testing before integration.
Another promising direction is to leverage iterative, feedback-driven refinement, where migration results are automatically validated and refined until they satisfy correctness, performance, and security constraints.
Such enhancements could significantly raise the success rate and robustness of existing migration techniques, enabling MBM to deliver more consistent and trustworthy maintenance outcomes at scale.

%% file: tex/4-source.tex
\section{Source Selection}

Choosing suitable sources for MBM faces three key challenges.
First, discovering relevant sources often relies on manual effort, with limited automation for finding forks, upstream/downstream projects, or similar systems.
Second, existing work focuses on code-level artifacts, ignoring other valuable migratable content such as usage data or issue reports.
Finally, filtering candidate commits or changes remains manual and costly, limiting scalability and introducing bias.

\subsection{C4. Uncovered Relevant Sources} Current MBM research relies on a narrow and manually selected set of data sources, focusing mostly on forks, upstream repositories, previous versions, and functionally similar projects. While these sources have proven useful for migration, they represent only a subset of the potential knowledge available in the broader software ecosystem. Moreover, even for well-known source types, identifying all relevant ones for a target project often depends on manual effort or prior domain knowledge. For example, Pan et al. ~\cite{Pan2024Patch} focus on Vim and Neovim and identify transferable patches by mining commit logs and manually inspecting candidate commits, which implicitly assumes that the selected patches are indeed worth migrating.
This strong preliminary assumption ensures that the migration data is directly reusable but skips deciding whether the new patch in Vim is actually relevant and should be migrated to Neovim, which is the crucial first step in real-world maintenance.
This limitation restricts the scalability and generalizability of MBM studies and hinders their application to real-world, large-scale software ecosystems. Future work should focus on automated discovery and ranking of relevant sources, reducing human effort, broadening the scope of tasks supported by MBM, and enabling more robust, ecosystem-wide maintenance strategies.

\textbf{O10. Broadening the Scope of Strongly Aligned Sources.} Current MBM research has primarily utilized open-source forks and upstream repositories. However, strictly aligned proprietary or ecosystem-specific sources remain underexplored.
One such source is \textit{vendor-maintained or enterprise editions}, which often backport security fixes and performance patches earlier than the community version. 
Systematically mining these changes and migrating them back to the open-source counterpart could accelerate vulnerability remediation and improve code quality for the wider ecosystem.  
Another promising direction is \textit{plugin and extension ecosystems}. Plugins frequently adapt to core API evolution faster than downstream clients, serving as rich references for API usage updates. 
Mining plugin adaptation commits could help automate API migration in other dependent projects and provide insights into common breakages caused by framework updates. 
Exploring these additional strongly aligned sources can enhance MBM's coverage and reduce redundant maintenance effort across ecosystems.

\textbf{O11. Automating the Discovery of Loosely Aligned Sources.}
Existing MBM research~\cite{Gao2024Test, Pan2024Patch} often relies on manually specifying which source repository to pair with a given target repository. 
However, identifying functionally similar projects that could serve as promising migration sources is itself time-consuming, error-prone, and requires significant domain expertise. 
For example, developers or researchers must manually browse ecosystems, inspect documentation, and compare codebases to locate candidate sources that share similar functionality, API usage, or architectural patterns. 
This manual process limits the scale of MBM studies and risks missing valuable sources that could improve migration quality. Recent advances in LLMs and repository-level code representation learning show strong potential for automating this step. 
LLMs can parse code, documentation, and commit history holistically, reason about functionality, and even summarize project intent in natural language. 
By leveraging these capabilities, future approaches could automatically retrieve and rank candidate source projects according to functional similarity, ecosystem relevance, and compatibility with the target. 
Integrating this into a recommendation pipeline could allow MBM systems to explore a much larger search space of candidate repositories, reducing human effort, improving reproducibility, and increasing the diversity of migration sources considered without manual intervention.

\textbf{O12. Leveraging Bidirectional Migration.} Most MBM research assumes a one-way flow of maintenance knowledge, migrating patches, tests, or exploits from a source project to a target project. However, once a mapping between source and target has been established, it can be reused to enable reverse migration, allowing both projects to continuously benefit from each other’s maintenance rather than keeping maintenance knowledge siloed. This concept has concrete applications. For example, in GUI testing, Fazzini et al.~\cite{Fazzini2019API} transfer test cases from iOS to Android apps, but the same widget and event mappings could also be used to migrate newly written Android tests back to iOS. In exploit migration, Zou et al.~\cite{zou2024syzbridge} move exploits from upstream kernels to downstream distributions, yet the reverse direction could help verify patch completeness or detect regressions in upstream. Similarly, when forks or downstream projects receive bug fixes, these fixes can be synchronized back upstream to reduce divergence and long-term maintenance debt. By extending migration pipelines, researchers can build symmetric maintenance workflows that continuously propagate knowledge between related projects or versions, with future work needed to tackle conflict resolution, deduplication, and consistency maintenance under concurrent evolution.

\textbf{O13. Toward a Unified Project–Source Map.}
Instead of repeatedly discovering and matching sources for each individual maintenance task, a promising direction is to build a comprehensive \textit{project–source knowledge graph} that encodes relationships such as forks, upstream–downstream dependencies, historical versions, cross-language siblings, and functional similarity. 
Such a database would enable researchers and practitioners to quickly retrieve all relevant sources for a given project and task, 
leveraging multiple sources simultaneously to improve migration accuracy and coverage.  
Moreover, for newly encountered maintenance tasks, this database would eliminate redundant source-discovery efforts and serve as a shared foundation for automated MBM pipelines, accelerating the adoption of MBM.

\subsection{C5. Limited Migratable Content} Current MBM research mainly focuses on code-level artifacts like patches, tests. However, reference projects often include other valuable resources that can provide rich maintenance insights. For example, user logs can be converted into realistic regression tests, and issue reports can offer hints for bug localization.
The challenge is not to migrate these raw data directly, but to turn them into useful guidance for existing maintenance tasks.
Future work should explore how to extract, filter, and align these signals to make them actionable for migration.

\textbf{O14. Leveraging Usage Data for Test Generation.} Real-world usage data offers a rich source of knowledge for test generation, especially for systems where manual test authoring is difficult or coverage is incomplete. For instance, in blockchain systems, replaying historical transactions can expose rare corner cases and capture stateful behaviors that would be hard to synthesize manually~\cite{Zhang2025chain}. Beyond blockchain, this opportunity generalizes to other domains, such as API invocation logs that can be abstracted into integration tests, GUI event traces that can be mapped to functional workflows and replayed on new app versions, and usage-based fault localization that leverages production traces and user interactions to reproduce failures. By transforming usage data into structured, replayable tests and migrating them across projects, MBM can provide realistic, high-value coverage and reduce the burden on developers to handcraft exhaustive test suites. 
Such tests can also capture rare corner cases and production-specific behaviors that are hard to anticipate manually.

\textbf{O15. Deriving Tests from Issue Reports.} Bug reports and issue trackers often include minimal reproducing examples, stack traces, input data, or detailed steps to reproduce a failure. These artifacts can serve as valuable raw material for generating executable regression tests. Migration-based approaches could systematically extract such reproducible scenarios, transform them into test cases, and adapt them for use in related projects or newer versions of the same project. For example, an issue containing a failing input and expected output can be converted into an automated unit test and replayed across multiple versions to verify whether the bug persists or has been fixed. Similarly, for projects in the same ecosystem, issue-derived tests can be migrated to detect whether similar defects exist elsewhere—enabling cross-project bug detection. Key research opportunities include automatically parsing issue descriptions, identifying reproducible code snippets or failing inputs, and synthesizing them into executable tests. Leveraging LLMs and program analysis techniques could improve the accuracy of issue-to-test conversion and enable their seamless adaptation across different codebases, thereby increasing regression test coverage and reducing the likelihood of bug recurrence.

\textbf{O16. Mining Issue Reports for Defect Discovery.} In the MBM paradigm, issue reports can be treated as rich sources of migratable knowledge. Rather than being static records of past failures, issues can be mined, abstracted, and transferred across projects or versions to proactively detect similar defects. For example, a concurrency deadlock reported in one project could be transformed into a symbolic execution condition or regression test and replayed on sibling forks or downstream versions that share the same concurrency primitives. This aligns with MBM’s two-step process: (1) matching issue patterns to corresponding locations in the target system (e.g., through call-graph matching or semantic similarity of stack traces) and (2) adapting them into tests. By migrating issue-derived knowledge, MBM can accelerate defect discovery in related systems and reduce redundant debugging efforts. Future research opportunities include automated generalization of issue-triggering conditions, cross-project matching of root-cause code regions, and adaptive replay techniques that account for environmental or version differences.

\subsection{C6. Lack of Automated Filtering.} After selecting a data source, researchers still face the challenge of identifying which artifacts within it are suitable for migration. Current MBM studies often rely on manual curation, such as handpicking candidate commits, patches, or test cases from forks and manually judging their transferability~\cite{Pan2024Patch, Gao2024Test}. This manual filtering is labor-intensive and introduces bias, as non-migratable artifacts may be included and waste computation, while valuable ones might be overlooked. Moreover, the proportion of truly migratable content within a source is typically small, make naive attempts at exhaustive migration inefficient and noisy. A key research challenge is therefore to develop automated filtering mechanisms that predict migratability before performing costly matching and adaptation. Promising directions include using static and dynamic analyses to pre-classify commits or patches, leveraging LLMs to semantically understand commit messages or diff context, and constructing benchmark datasets with both positive and negative examples to evaluate filtering techniques. Automating this step would substantially improve the precision and cost-effectiveness of MBM pipelines.

\textbf{O17. Target Operation Identification.} Even after selecting a suitable data source, MBM must determine which operations are actually relevant to the target maintenance task. Current studies often rely on manual inspection or coarse-grained heuristics to filter candidate operations, which is inefficient and prone to omissions. Future MBM research could leverage program analysis, information retrieval, and LLM-based reasoning to automatically infer the intent of each candidate operation and align it with the maintenance objective. For example, commit messages and code diffs could be semantically classified to detect bug-fix commits or performance optimizations, while test cases could be ranked by their coverage relevance to the target system. Building automated pipelines for target operation identification would significantly reduce human effort, improve migration precision, and enable MBM to scale to large, noisy repositories.

\textbf{O18. Migration Knowledge Reuse.} When related projects have already undergone similar migration efforts, the resulting mappings, transformed patches, or migrated tests can serve as valuable prior knowledge. Instead of re-performing the entire matching and adaptation process from scratch, MBM pipelines could reuse these previously derived migration artifacts to accelerate maintenance for new target projects. For example, if a patch has been ported between two forks, the same transformation pattern can be applied to other forks with similar code context. Likewise, established API mappings or GUI widget alignments could be stored in a shared knowledge base and automatically applied in future migrations. Systematically capturing, indexing, and reusing migration knowledge would both improve efficiency and enhance the consistency across projects.

%% file: tex/5-adaption.tex
\section{Source Adjustment}

Source adjustment is the stage that differentiates MBM from purely data-driven or history-based approaches, as it requires adapting retrieved artifacts to fit the target system.
However, this stage remains one of the most technically challenging parts of the MBM pipeline.
Key challenges arise from \ding{182} the limited accuracy and generalizability of current matching techniques, which often fail when the source and target diverge semantically or structurally,
\ding{183} the low success rate of migration in complex scenarios, such as long dependency chains, cross-language contexts, or multi-project environments, and
\ding{184} the lack of interpretability in migration operations, which reduces trust and makes debugging or auditing difficult.
Addressing these challenges is essential for making MBM robust, reliable, and acceptable in real-world software maintenance practice.

\subsection{C7. Insufficiently Accurate and Generalizable Matching} 
Current matching approaches in MBM often rely on shallow features such as method or class names, or token-level similarity after simple preprocessing.
While these strategies work reasonably well for simple code changes such as refactorings in sources with high structural similarity, such as forks or earlier versions, they struggle when code structure diverges due to cross-language differences or architectural evolution.
Even methods leveraging embeddings are still largely text-driven, lacking deeper semantic understanding or runtime context.
This leads to lower matching precision and recall, and poor generalization across projects, languages, or frameworks, ultimately reducing the reliability of the migration process.

\textbf{O19. Rich, Multi-Modal Matching Scopes.}
Current matching methods often rely only on surface text features such as method names or code-level token overlap, which may miss deeper structural or semantic relationships. 
Future research should broaden the information leveraged for matching by incorporating diverse signals such as ASTs, control/data-flow graphs, type hierarchies, and runtime traces. 
These complementary views can be fused into a unified representation or learned embedding space, allowing matching to consider structure, behavior, and textual semantics jointly. 
This richer signal space would improve robustness against naming inconsistencies and enable more generalizable matching across projects, versions, and languages. 
In particular, multi-modal matching is critical for complex scenarios such as cross-language migration or highly refactored codebases, where textual clues alone are insufficient. 

\textbf{O20. Beyond Textual Similarity.}
Beyond computing similarity, future approaches should validate candidate matches through semantic and behavioral equivalence checks.
Static analysis can verify type compatibility, call graph integrity, and API contract compliance, while dynamic analysis can replay source tests or usage traces on candidate targets to confirm behavioral parity.
Recent advances in LLMs provide a new opportunity to enhance matching by reasoning over code intent, comments, and surrounding context rather than relying solely on surface similarity. LLMs can infer latent correspondences between functions or APIs even across languages and propose candidate mappings with explanatory rationales.
However, LLM predictions can be noisy or overconfident, so they should be integrated into a verification pipeline. Compiler checks, static type analysis, and differential testing can filter out hallucinated or semantically unsafe matches.

\textbf{O21. Multi-Source Matching and Consensus Building.}
A promising direction is to perform matching not from a single source, but across multiple related sources simultaneously. In many ecosystems, the same functionality may appear in several forks, downstream distributions, and cross-language implementations. Instead of relying on one-to-one mapping, future approaches could aggregate candidate matches from multiple sources and reconcile them to reach a consensus. For example, if a function appears with consistent naming and similar call patterns in three forks but diverges in a fourth, this information can be used to raise confidence in the majority mapping or flag potential divergences for developer review. Rather than focusing on single-source similarity, MBM techniques could combine structural, historical, and behavioral evidence from multiple sources to build a stronger case for each match. Such consensus-driven matching would reduce the noise and bias of any individual source, improve robustness in structurally divergent scenarios, and give developers more reliable migration candidates to work with.

\subsection{C8. Low-Efficiency Migration for Complex Scenarios} 
A recurring challenge for MBM is that migration techniques achieve high precision and recall primarily in structurally simple cases but struggle with complex, real-world maintenance scenarios.
When the source and target systems diverge significantly, current approaches often fail to produce usable results.
For example, patch porting studies report success rates as low as 40–45\% on hard forks~\cite{Pan2024Patch}, where semantic drift and code divergence accumulate over time. GUI test migration across platforms may achieve high recall but still suffer from precision loss due to mismatched widgets, timing issues, or missing event handlers.
This gap highlights the need for methods that are resilient to heterogeneity, robust across ecosystems, and capable of handling the long tail of migration cases that are most common in practice.

\textbf{O22. LLM-Augmented Migration Components.}
Current evidence suggests that LLMs still struggle to directly perform unseen maintenance tasks in an end-to-end manner~\cite{Pan2024Patch}, often producing incomplete or incorrect migrations in complex scenarios. A promising opportunity is to integrate LLMs as intelligent augmentations for specific MBM components.
For example, LLMs can replace simplistic string-based matching with semantic-aware matching, using their ability to reason about API intent, class roles, or even refactoring patterns across projects. They can enhance test generation during migration, offering a principled alternative to mutation-based tools like EvoSuite in migration tasks~\cite{Chen2024Exploit}. In test migration~\cite{Gao2024Test}, LLMs can replace brittle, hand-crafted formal rules with learned, data-driven translators that align GUI elements and assertions across platforms and versions. Finally, hybrid approaches could couple LLM reasoning with fuzzers, where the LLM generates diverse, high-value seeds and the fuzzer explores the search space more effectively, achieving deeper coverage of complex migration scenarios.
By modularizing LLM contributions across MBM pipelines, researchers can systematically improve accuracy, reduce human intervention, and create a feedback loop where validated results are used to further refine the process.

\textbf{O23. Cross-Language Consistency Guarantees.}
Cross-language migration introduces semantic gaps that go beyond syntax or AST-level structure, making simple name matching or embedding similarity insufficient.
In multilingual projects, such as those with Java/C mirror implementations or mixed C/C++/Python stacks, developers must ensure that fixes, patches, and refactorings preserve behavioral equivalence across all languages.
Promising directions include constructing language-agnostic intermediate representations that capture program intent and validating consistency through semantic equivalence checking or property-based testing.
For example, bug fixes applied to a Java implementation could be validated against a C implementation using shared behavioral specifications or automatically derived regression tests.
This would prevent silent divergence between implementations, a risk that can accumulate over time and lead to subtle inconsistencies in production systems.

\textbf{O24. Handling Long Call Graph in Upstream–Downstream Migration.}
A major limitation of current MBM techniques is their inability to handle cases where maintenance effects propagate through long and complex upstream–downstream call chains. 
For instance, in exploit migration tasks, a vulnerability trigger in an upstream library may pass through multiple layers of wrappers and adapters before reaching the vulnerable sink in downstream applications. 
Existing methods often fail in such cases because they cannot synthesize composite triggering conditions that traverse these multi-hop call paths, leading to incomplete or failed migrations~\cite{Chen2024Exploit}. 
Future work should explore hybrid static–dynamic approaches that recover full inter-procedural call chains, generate combined preconditions for deep call sequences, and dynamically replay them in downstream contexts. 
By reconstructing the full propagation chain, MBM systems could reproduce vulnerability-triggering behaviors with higher fidelity and improve downstream impact assessment. 
Such techniques would reduce under-reporting of affected systems and significantly improve migration robustness in complex software ecosystems.

\subsection{C9. Lack of Interpretability of Migration Operations} A persistent challenge in source adjustment is the lack of interpretability and transparency in migration decisions.
Most current MBM methods output the final migrated artifact without explaining why a specific source was selected, which transformation rules were applied, or what potential side effects might occur.
This black-box nature makes it hard for developers to trust migration results and complicates debugging when the migrated result fails.
Interpretable migration operations are crucial to support developer adoption, facilitate review processes, and ensure that automated maintenance aligns with project-specific requirements.

\textbf{O25. Justification of Source Selection.} 
A major barrier to trust is the lack of transparency in source selection. Developers often struggle to understand why a specific reference (e.g., a patch or test case) was chosen to guide the migration.
Future research should focus on explaining the relevance and feasibility of the selected sources. Instead of presenting a result in isolation, MBM systems could provide a relevance report that highlights the similarities between the project under maintenance and the retrieved reference. For example, in patch migration, the tool could visualize the alignment of stack traces or buggy code patterns, demonstrating that the selected reference addresses the exact same root cause. By making the source selection transparent, developers can verify that the migration is built on a valid and applicable foundation.

\textbf{O26. Rationalization of Adaptation Logic.} Even with a valid source, the adaptation process remains a black box. A critical opportunity lies in explaining the transformation logic used to adapt the source to target contexts. Developers need to understand why specific edits were applied, such as why a variable was renamed, why a conditional check was added, or why a function call was invoked. Future MBM systems could generate natural language rationales for each edit operation (e.g., Renamed variable x to match the target scope). This level of detail moves beyond simple code diffs, helping developers distinguish between necessary structural adaptations and potential model hallucinations, thereby streamlining code review and debugging.

%% file: tex/6-evaluation.tex
\section{Result Evaluation}
\label{Evaluate}

Evaluating the correctness and impact of MBM tasks poses three major challenges. 
First, most studies rely on manually annotated ground truth, which is infeasible in real-world maintenance, where annotations are missing, and migration outcomes must be validated automatically. 
Second, existing metrics rarely assess the broader impact of migration, such as whether the introduced changes preserve original functionality or cause regressions. 
Finally, evaluation datasets are often small, task-specific, and lack reusable oracles, limiting reproducibility and cross-method comparisons.

\subsection{C10. Lack of Automated and Comprehensive Metrics.} Current MBM research largely relies on human-in-the-loop evaluation to assess migration correctness, such as human inspection of patch equivalence~\cite{Pan2024Patch} or manual validation of migrated tests~\cite{Zhang2024GUI}. 
These approaches are costly, making them impractical for real-world maintenance. 
Moreover, existing research sometimes fails to assess whether the maintenance objective has been truly achieved, for example, by examining code similarity. 
This gap prevents researchers from fairly comparing MBM methods across tasks and hinders developers from trusting migration results in production.


\textbf{O27. Self-Verifying Maintenance Based on Migration.}
\dx{A critical barrier to the adoption of MBM is the reliance on human validation. Current evaluations largely depend on manual review, as there is no automatic guarantee that a migrated artifact works as intended. The unique opportunity of MBM lies in the fact that the maintenance goal is already verified in the source project. Therefore, the source serves not just as a code template, but as a Behavioral Oracle. Future research should focus on transferring verification assets alongside the production code. This allows MBM to establish a verification-driven feedback loop, enabling the system to iteratively repair and refine the code after migration without human intervention.}

\textbf{O28. Metrics for Migration Validity.}  
\dx{The migrated artifact must be semantically equivalent to the source (to preserve the logic) but syntactically different (to fit the target context). To achieve this, future evaluations should focus on two metrics: Semantic Consistency and Adaptation Accuracy. However, existing studies predominantly overlook this duality, relying instead solely on textual similarity (e.g., BLEU) and simple compilation checks. These metrics are insufficient as they cannot determine whether a migration is both logically sound and contextually matched. Future research should develop metrics specific to the migration process. First, Semantic Consistency metrics are needed to verify that the core logic remains invariant after migration. For instance, in test migration, instead of just checking execution success, one should measure whether the migrated test in the target triggers the expected execution path or state change as it did in the source. Second, Adaptation Accuracy metrics should quantify how well the migrated artifact integrates into the target system. A successful migration should not only function in isolation but also maintain a zero-regression baseline in the existing verification workflow. For example, if a migrated patch causes failures in existing tests, it indicates a failure to adapt to the context. }

\subsection{C11. Insufficient Functional Impact Assessment.}
A major challenge in result evaluation is the lack of systematic assessment of how MBM operations affect the overall functionality of the target system. 
Current approaches often stop at verifying that the migrated artifact compiles or passes basic tests, but do not comprehensively evaluate whether the migration preserves intended behavior. 
For instance, a patch or API migration may fix a fault but inadvertently break unrelated features. 
Robust functional impact assessments are required to ensure that post-migration systems maintain correctness, performance, and security properties.

\textbf{O29. Automated Differential Impact Assessment.}  
A promising direction is to establish differential testing as the default paradigm for evaluating MBM outcomes. 
Instead of merely verifying that migrated code compiles or runs, future approaches could systematically compare the behavior of pre- and post-migration systems under a shared input space. 
To maximize coverage, this input space can be enriched by multiple complementary sources. 
For well-maintained source projects, their existing regression tests can be migrated to the target project to quickly establish a baseline of behavioral checks and catch regressions early. 
For projects that lack adequate tests, automatically generated tests or fuzzing can fill critical coverage gaps and exercise rarely tested code paths. 
Finally, real-world usage traces together with tests migrated from source projects can be leveraged to validate behavior under realistic workloads and corner cases, ensuring that the migrated system functions correctly in practice.
The goal is to automatically detect semantic divergences introduced by migration without heavy manual inspection. 
A unified differential evaluation pipeline would provide a principled way to quantify the functional impact of MBM operations and serve as a robust safeguard before deploying migration results into production.

\textbf{O30. Change-Aware Oracle Generation.}
Migration typically modifies only a limited portion of the target system, yet validating correctness often requires re-running large test suites or generating many redundant tests.
Change-aware test generation focuses specifically on the regions impacted by migration operations, providing a more efficient and targeted verification process.
This approach could combine program-difference analysis to locate code regions affected by migrated patches or API updates, dependency slicing to trace transitive impact on related modules, and automated synthesis of focused regression tests for these regions.
MBM offers unique opportunities to guide this process: the migration context and source project’s tests can reveal which functionalities are most likely to be affected, enabling prioritization of test generation efforts.
Future work could integrate static and dynamic impact analysis with modern test generation frameworks and leverage LLMs to synthesize human-readable oracles that explain behavioral differences, improving both test precision and interpretability.
Such approaches could significantly reduce validation costs while improving the likelihood of catching migration-induced regressions.

\textbf{O31. LLM-as-Judge for Functional Impact Assessment.}
The rapid development of LLMs offers an intriguing opportunity to assess whether migration operations preserve the intended functionality of the target system.
Rather than merely checking compilation or running a limited test suite, LLMs can be prompted with both the pre- and post-migration code, commit messages, and contextual information to reason about whether the migration might introduce behavioral regressions or break unrelated features. This approach could help project maintainers answer a key question: Does this migrated change maintain correctness and avoid unintended side effects?
For example, given a patch migration, an LLM could summarize the semantic difference between source and target patches and flag suspicious divergences.
Similarly, for API or configuration migration, the LLM could reason about whether the migrated call sequence still satisfies preconditions or preserves invariants described in the documentation. However, MBM-specific challenges remain: LLMs may hallucinate or miss subtle semantic issues, and their assessments must be trustworthy, reproducible, and explainable enough to support real-world maintenance decisions.
Future work could enhance LLMs for this task by fine-tuning them on migration datasets labeled as safe vs. breaking, and combining their judgments with differential testing or static analysis signals to improve reliability.

\subsection{C12. Scarcity of Shared, Reusable Datasets.}
Another key challenge in result evaluation is the lack of publicly available, well-constructed datasets for MBM research. 
Most existing studies build datasets tailored to a specific task, often focusing on a narrow set of projects or manually curated examples. 
This fragmentation hinders reproducibility, makes it difficult to perform fair cross-method comparisons, and prevents cumulative progress across studies. 
Moreover, the absence of standardized benchmarks means that researchers must repeatedly invest effort in data collection and annotation, increasing the barrier to entry for new work. 
A concerted effort is needed to create large-scale and automatically maintainable MBM benchmarks that not only provide ground-truth mappings but also include automated oracles for verifying whether the migrated maintenance operations achieve their intended goals, as well as whether the original system functionality remains intact after migration.

\textbf{O32. Comprehensive, Function-Aware MBM Benchmarks.} 
A promising opportunity lies in the construction of comprehensive, function-aware benchmark datasets for MBM. 
Such datasets could systematically collect functionally similar modules across projects and ecosystems, along with their associated commits, tests, issues, and real-world usage data. 
These resources would enable rigorous evaluation of migration techniques by testing whether a method can successfully transfer maintenance artifacts across multiple implementations of the same functionality. 
For example, researchers could evaluate whether a test case for a JSON parsing library in Java can be migrated to other Java JSON libraries (e.g., Gson, Jackson), or even to Python JSON libraries, thereby assessing both intra-language and cross-language migration capabilities. 
Such benchmarks would expose weaknesses in existing approaches, promote fair comparisons across methods, and accelerate progress toward robust and generalizable MBM techniques.

\textbf{O33. Mining Real-World Migration Datasets.}  
A promising direction is to systematically mine and curate migration phenomena already present in well-maintained real projects. 
Examples include cherry-picked commits across LTS branches, patch porting between forks, and test reuse across releases or sibling systems. 
Such data are valuable because they capture developer-validated maintenance operations. 
Automatically collecting these instances at scale would provide an authentic ground truth for benchmarking MBM techniques and studying migration patterns. Beyond serving as a static benchmark, these mined migration instances can enable \emph{differential evaluation} of MBM methods. By comparing automatically generated migrations against developer-approved ones, researchers can quantify correctness, completeness, and adaptation effort. 
This would allow the community to measure not only whether a migration is possible, but also how close automated results are to real-world maintenance practice, ultimately driving the refinement of MBM techniques toward production-level reliability.

\textbf{O34. Systematic Assessment of Existing MBM Tools.}  
While real-world migration datasets can serve as benchmarks, there is also a need for comprehensive studies that answer the question: \textit{``How far are we?''} in MBM research. 
Rather than focusing on isolated case studies or single maintenance tasks, future work should conduct large-scale, systematic evaluations of state-of-the-art MBM approaches across a curated and diverse suite of migration scenarios. 
Such assessments should quantify success rates, measure cost savings (e.g., reduced developer effort or build failures), and analyze common failure cases to expose bottlenecks in matching, adaptation, and validation.  
This form of meta-analysis would not only highlight where current techniques excel and where they break down, but also reveal which kinds of migration tasks remain unsolved, thus guiding research investments toward the most impactful gaps.  

%% file: tex/7-perspective.tex
\section{Perspective}
\label{Perspective}

Software maintenance is the backbone of long-term software sustainability, ensuring reliability, security, and usability even as systems grow and evolve.
As modern software ecosystems become increasingly large, interconnected, and fast-evolving, we believe that \textbf{Migration-Based Maintenance} will play a pivotal role in enabling scalable, automated, and trustworthy maintenance.
Looking ahead, we see several key shifts in the role and scope of MBM.

First, MBM will broaden its task coverage.
Rather than focusing narrowly on patch porting, test reuse, or API updating, MBM must expand to encompass performance optimization, configuration evolution, and documentation synchronization.
The rise of emerging domains, such as deep learning frameworks and smart contracts, presents new opportunities for MBM to address the unique maintenance challenges of these complex environments, thereby reducing engineering costs and ensuring system reliability.

Second, methodological advances will reshape how MBM is performed.
Progress in LLMs is expected to enable more accurate source–target matching, semantic adaptation of migrated artifacts, and predictive assessment of migration feasibility.
We envision hybrid pipelines in which LLMs leverage semantic reasoning to identify data sources and perform migrations, while static and dynamic analyses validate correctness through compilation checks and differential testing.
Over time, ecosystem-scale project-source knowledge graphs could enrich the data foundation of MBM, facilitating more comprehensive and robust migrations.

Third, MBM must be seamlessly integrated into CI/CD pipelines to automatically trigger existing tests. The results from these tests can serve as a feedback loop, helping to refine the migration process. Furthermore, to foster developer trust, it is essential to accompany automatically generated operations with explainable rationales, ensuring that developers can understand and approve the migration decisions.
Such traceable and interpretable pipelines will be crucial for enabling maintainers to adopt automated changes with confidence.
Multi-dimensional evaluation metrics that jointly consider correctness, performance, security, and maintainability will further ensure that automated migrations align with project goals.

Finally, we see an increasing need for community-wide benchmarks and systematic evaluations.
Shared datasets and comprehensive assessments across diverse ecosystems will allow the research community to measure how far current MBM techniques have progressed, identify persistent failure cases, and prioritize the most impactful research directions.

Overall, we are optimistic that in the coming decade, MBM will become a cornerstone of sustainable software engineering practice.
Its success will be measured not only by technical accuracy but also by its ability to make automated maintenance less burdensome, more reliable, and more transparent.

%% file: main.bib
@article{Tang2025mirror,
author = {Tang, Ye and Chen, Honghao and He, Zhixing and Zhong, Hao},
title = {Understanding Mirror Bugs in Multiple-Language Projects},
year = {2025},
publisher = {Association for Computing Machinery},
address = {New York, NY, USA},
issn = {1049-331X},
url = {https://doi.org/10.1145/3727145},
doi = {10.1145/3727145},
note = {Just Accepted},
journal = {ACM Trans. Softw. Eng. Methodol.},
month = mar
}

@misc{Toolkit,
    title = {HTTP Toolkit Configuration Migration},
    url = {https://httptoolkit.com/blog/migrating-javascript-from-travis-to-github-actions/},
    author = {Tim Perry},
}

@inproceedings{PICOC,
author = {Kitchenham, Barbara A.},
title = {Systematic review in software engineering: where we are and where we should be going},
year = {2012},
isbn = {9781450315098},
publisher = {Association for Computing Machinery},
address = {New York, NY, USA},
url = {https://doi.org/10.1145/2372233.2372235},
doi = {10.1145/2372233.2372235},
booktitle = {Proceedings of the 2nd International Workshop on Evidential Assessment of Software Technologies},
pages = {1–2},
numpages = {2},
location = {Lund, Sweden},
series = {EAST '12}
}

@misc{pythonIssue,
    title = {Issue: Python -3 Documentation is Outdated},
    url = {https://bugs.python.org/issue21434},
    author = {Python Software Foundation},
}

@misc{OpenSourceGo,
    title = {Open-Source Go Configuration Migration},
    url = {https://markphelps.me/posts/migrating-from-travis-to-github-actions/},
    author = {Mark Phelps},
}

@inproceedings{Yin2025Repair,
author = {Yin, Xin and Ni, Chao and Wang, Shaohua and Li, Zhenhao and Zeng, Limin and Yang, Xiaohu},
title = {ThinkRepair: Self-Directed Automated Program Repair},
year = {2024},
isbn = {9798400706127},
publisher = {Association for Computing Machinery},
address = {New York, NY, USA},
url = {https://doi.org/10.1145/3650212.3680359},
doi = {10.1145/3650212.3680359},
booktitle = {Proceedings of the 33rd ACM SIGSOFT International Symposium on Software Testing and Analysis},
pages = {1274–1286},
numpages = {13},
location = {Vienna, Austria},
series = {ISSTA 2024}
}

@article{Guan2025DLbug,
author = {Guan, Hao and Bai, Guangdong and Liu, Yepang},
title = {CrossProbe: LLM-Empowered Cross-Project Bug Detection for Deep Learning Frameworks},
year = {2025},
issue_date = {July 2025},
publisher = {Association for Computing Machinery},
address = {New York, NY, USA},
volume = {2},
number = {ISSTA},
url = {https://doi.org/10.1145/3728984},
doi = {10.1145/3728984},
journal = {Proc. ACM Softw. Eng.},
month = jun,
articleno = {ISSTA107},
numpages = {23}
}

@INPROCEEDINGS{Shen2025DL,
  author={Shen, Qingchao and Tian, Yongqiang and Ma, Haoyang and Chen, Junjie and Huang, Lili and Fu, Ruifeng and Cheung, Shing-Chi and Wang, Zan},
  booktitle={2025 IEEE/ACM 47th International Conference on Software Engineering (ICSE)}, 
  title={A Tale of Two DL Cities: When Library Tests Meet Compiler}, 
  year={2025},
  volume={},
  number={},
  pages={2201-2212},
  doi={10.1109/ICSE55347.2025.00025}}

@inproceedings{Zhang2025chain,
author = {Zhang, Jiashuo and Chen, Jiachi and Grundy, John and Gao, Jianbo and Wang, Yanlin and Chen, Ting and Guan, Zhi and Chen, Zhong},
title = {Automated Test Generation for Smart Contracts via On-Chain Test Case Augmentation and Migration},
year = {2025},
isbn = {9798331505691},
publisher = {IEEE Press},
url = {https://doi.org/10.1109/ICSE55347.2025.00096},
doi = {10.1109/ICSE55347.2025.00096},
booktitle = {Proceedings of the IEEE/ACM 47th International Conference on Software Engineering},
pages = {1947–1959},
numpages = {13},
location = {Ottawa, Ontario, Canada},
series = {ICSE '25}
}

@INPROCEEDINGS{Dorninger2017Documentation,
  author={Dorninger, Bernhard and Moser, Michael and Pichler, Josef},
  booktitle={2017 IEEE 24th International Conference on Software Analysis, Evolution and Reengineering (SANER)}, 
  title={Multi-language re-documentation to support a COBOL to Java migration project}, 
  year={2017},
  volume={},
  number={},
  pages={536-540},
  doi={10.1109/SANER.2017.7884669}}

@ARTICLE{LeGoues2012APR,
  author={Le Goues, Claire and Nguyen, ThanhVu and Forrest, Stephanie and Weimer, Westley},
  journal={IEEE Transactions on Software Engineering}, 
  title={GenProg: A Generic Method for Automatic Software Repair}, 
  year={2012},
  volume={38},
  number={1},
  pages={54-72},
  doi={10.1109/TSE.2011.104}}

@inproceedings{Jiang2018APR,
author = {Jiang, Jiajun and Xiong, Yingfei and Zhang, Hongyu and Gao, Qing and Chen, Xiangqun},
title = {Shaping program repair space with existing patches and similar code},
year = {2018},
isbn = {9781450356992},
publisher = {Association for Computing Machinery},
address = {New York, NY, USA},
url = {https://doi.org/10.1145/3213846.3213871},
doi = {10.1145/3213846.3213871},
booktitle = {Proceedings of the 27th ACM SIGSOFT International Symposium on Software Testing and Analysis},
pages = {298–309},
numpages = {12},
location = {Amsterdam, Netherlands},
series = {ISSTA 2018}
}

@article{Xuan2017APR,
author = {Xuan, Jifeng and Martinez, Matias and DeMarco, Favio and Clement, Maxime and Marcote, Sebastian Lamelas and Durieux, Thomas and Le Berre, Daniel and Monperrus, Martin},
title = {Nopol: Automatic Repair of Conditional Statement Bugs in Java Programs},
year = {2017},
issue_date = {January 2017},
publisher = {IEEE Press},
volume = {43},
number = {1},
issn = {0098-5589},
url = {https://doi.org/10.1109/TSE.2016.2560811},
doi = {10.1109/TSE.2016.2560811},
journal = {IEEE Trans. Softw. Eng.},
month = jan,
pages = {34–55},
numpages = {22}
}

@INPROCEEDINGS{Nguyen2013APR,
  author={Nguyen, Hoang Duong Thien and Qi, Dawei and Roychoudhury, Abhik and Chandra, Satish},
  booktitle={2013 35th International Conference on Software Engineering (ICSE)}, 
  title={SemFix: Program repair via semantic analysis}, 
  year={2013},
  volume={},
  number={},
  pages={772-781},
  doi={10.1109/ICSE.2013.6606623}}

@article{Koyuncu2020APR,
author = {Koyuncu, Anil and Liu, Kui and Bissyand\'{e}, Tegawend\'{e} F. and Kim, Dongsun and Klein, Jacques and Monperrus, Martin and Le Traon, Yves},
title = {FixMiner: Mining relevant fix patterns for automated program repair},
year = {2020},
issue_date = {May 2020},
publisher = {Kluwer Academic Publishers},
address = {USA},
volume = {25},
number = {3},
issn = {1382-3256},
url = {https://doi.org/10.1007/s10664-019-09780-z},
doi = {10.1007/s10664-019-09780-z},
journal = {Empirical Softw. Engg.},
month = may,
pages = {1980–2024},
numpages = {45}
}

@inproceedings{Xia2022APR,
author = {Xia, Chunqiu Steven and Zhang, Lingming},
title = {Less training, more repairing please: revisiting automated program repair via zero-shot learning},
year = {2022},
isbn = {9781450394130},
publisher = {Association for Computing Machinery},
address = {New York, NY, USA},
url = {https://doi.org/10.1145/3540250.3549101},
doi = {10.1145/3540250.3549101},
booktitle = {Proceedings of the 30th ACM Joint European Software Engineering Conference and Symposium on the Foundations of Software Engineering},
pages = {959–971},
numpages = {13},
location = {Singapore, Singapore},
series = {ESEC/FSE 2022}
}

@ARTICLE{Chen2021APR,
  author={Chen, Zimin and Kommrusch, Steve and Tufano, Michele and Pouchet, Louis-Noël and Poshyvanyk, Denys and Monperrus, Martin},
  journal={IEEE Transactions on Software Engineering}, 
  title={SequenceR: Sequence-to-Sequence Learning for End-to-End Program Repair}, 
  year={2021},
  volume={47},
  number={9},
  pages={1943-1959},
  doi={10.1109/TSE.2019.2940179}}

@inproceedings{Liu2019APR,
author = {Liu, Kui and Koyuncu, Anil and Kim, Dongsun and Bissyand\'{e}, Tegawend\'{e} F.},
title = {TBar: revisiting template-based automated program repair},
year = {2019},
isbn = {9781450362245},
publisher = {Association for Computing Machinery},
address = {New York, NY, USA},
url = {https://doi.org/10.1145/3293882.3330577},
doi = {10.1145/3293882.3330577},
booktitle = {Proceedings of the 28th ACM SIGSOFT International Symposium on Software Testing and Analysis},
pages = {31–42},
numpages = {12},
location = {Beijing, China},
series = {ISSTA 2019}
}

@article{Huang2024APR,
author = {Huang, Kai and Xu, Zhengzi and Yang, Su and Sun, Hongyu and Li, Xuejun and Yan, Zheng and Zhang, Yuqing},
title = {Evolving Paradigms in Automated Program Repair: Taxonomy, Challenges, and Opportunities},
year = {2024},
issue_date = {February 2025},
publisher = {Association for Computing Machinery},
address = {New York, NY, USA},
volume = {57},
number = {2},
issn = {0360-0300},
url = {https://doi.org/10.1145/3696450},
doi = {10.1145/3696450},
journal = {ACM Comput. Surv.},
month = oct,
articleno = {36},
numpages = {43}
}

@techreport{ISO14764,
  author      = {International Organization for Standardization},
  title       = {{ISO/IEC/IEEE 14764:2022}: Software engineering — Software life cycle processes — Maintenance},
  institution = {International Organization for Standardization},
  year        = {2022},
  type        = {Standard},
  url         = {https://www.iso.org/standard/80710.html}
}

@inproceedings{2008migration,
author = {Sch\"{a}fer, Thorsten and Jonas, Jan and Mezini, Mira},
title = {Mining framework usage changes from instantiation code},
year = {2008},
isbn = {9781605580791},
publisher = {Association for Computing Machinery},
address = {New York, NY, USA},
url = {https://doi.org/10.1145/1368088.1368153},
doi = {10.1145/1368088.1368153},
booktitle = {Proceedings of the 30th International Conference on Software Engineering},
pages = {471–480},
numpages = {10},
location = {Leipzig, Germany},
series = {ICSE '08}
}

@INPROCEEDINGS{Mukelabai2021ForkTest,
  author={Mukelabai, Mukelabai and Berger, Thorsten and Borba, Paulo},
  booktitle={2021 IEEE/ACM 43rd International Conference on Software Engineering: New Ideas and Emerging Results (ICSE-NIER)}, 
  title={Semi-Automated Test-Case Propagation in Fork Ecosystems}, 
  year={2021},
  volume={},
  number={},
  pages={46-50},
  doi={10.1109/ICSE-NIER52604.2021.00018}}

@inproceedings{Almeida2024Oneshot,
author = {Almeida, Aylton and Xavier, Laerte and Valente, Marco Tulio},
title = {Automatic Library Migration Using Large Language Models: First Results},
year = {2024},
isbn = {9798400710476},
publisher = {Association for Computing Machinery},
address = {New York, NY, USA},
url = {https://doi.org/10.1145/3674805.3690746},
doi = {10.1145/3674805.3690746},
booktitle = {Proceedings of the 18th ACM/IEEE International Symposium on Empirical Software Engineering and Measurement},
pages = {427–433},
numpages = {7},
location = {Barcelona, Spain},
series = {ESEM '24}
}

@article{Saboor2025repair,
author = {Saboor Yaraghi, Ahmadreza and Holden, Darren and Kahani, Nafiseh and Briand, Lionel},
title = {Automated Test Case Repair Using Language Models},
year = {2025},
issue_date = {April 2025},
publisher = {IEEE Press},
volume = {51},
number = {4},
issn = {0098-5589},
url = {https://doi.org/10.1109/TSE.2025.3541166},
doi = {10.1109/TSE.2025.3541166},
journal = {IEEE Trans. Softw. Eng.},
month = feb,
pages = {1104–1133},
numpages = {30}
}

@article{Elder2024Exploit,
author = {Elder, Sarah and Rahman, Md Rayhanur and Fringer, Gage and Kapoor, Kunal and Williams, Laurie},
title = {A Survey on Software Vulnerability Exploitability Assessment},
year = {2024},
issue_date = {August 2024},
publisher = {Association for Computing Machinery},
address = {New York, NY, USA},
volume = {56},
number = {8},
issn = {0360-0300},
url = {https://doi.org/10.1145/3648610},
doi = {10.1145/3648610},
journal = {ACM Comput. Surv.},
month = apr,
articleno = {205},
numpages = {41}
}

@misc{KernelBench,
      title={KernelBench: Can LLMs Write Efficient GPU Kernels?}, 
      author={Anne Ouyang and Simon Guo and Simran Arora and Alex L. Zhang and William Hu and Christopher Ré and Azalia Mirhoseini},
      year={2025},
      eprint={2502.10517},
      archivePrefix={arXiv},
      primaryClass={cs.LG},
      url={https://arxiv.org/abs/2502.10517}, 
}

@misc{CIgrate,
      title={CIgrate: Automating CI Service Migration with Large Language Models}, 
      author={Md Nazmul Hossain and Taher A. Ghaleb},
      year={2025},
      eprint={2507.20402},
      archivePrefix={arXiv},
      primaryClass={cs.SE},
      url={https://arxiv.org/abs/2507.20402}, 
}

@book{Kitchenham2015Review,
author = {Kitchenham, Barbara Ann and Budgen, David and Brereton, Pearl},
title = {Evidence-Based Software Engineering and Systematic Reviews},
year = {2015},
isbn = {1482228653},
publisher = {Chapman \& Hall/CRC}
}

@article{Zhou2024Repair,
author = {Zhou, Xin and Cao, Sicong and Sun, Xiaobing and Lo, David},
title = {Large Language Model for Vulnerability Detection and Repair: Literature Review and the Road Ahead},
year = {2024},
publisher = {Association for Computing Machinery},
address = {New York, NY, USA},
issn = {1049-331X},
url = {https://doi.org/10.1145/3708522},
doi = {10.1145/3708522},
note = {Just Accepted},
journal = {ACM Trans. Softw. Eng. Methodol.},
month = dec
}

@inproceedings{Chen2020API,
author = {Chen, Chunyang},
title = {SimilarAPI: mining analogical APIs for library migration},
year = {2020},
isbn = {9781450371223},
publisher = {Association for Computing Machinery},
address = {New York, NY, USA},
url = {https://doi.org/10.1145/3377812.3382140},
doi = {10.1145/3377812.3382140},
booktitle = {Proceedings of the ACM/IEEE 42nd International Conference on Software Engineering: Companion Proceedings},
pages = {37–40},
numpages = {4},
location = {Seoul, South Korea},
series = {ICSE '20}
}

@inproceedings{Collie2021API,
author = {Collie, Bruce and Ginsbach, Philip and Woodruff, Jackson and Rajan, Ajitha and O'Boyle, Michael F. P.},
title = {M3: semantic API migrations},
year = {2021},
isbn = {9781450367684},
publisher = {Association for Computing Machinery},
address = {New York, NY, USA},
url = {https://doi.org/10.1145/3324884.3416618},
doi = {10.1145/3324884.3416618},
booktitle = {Proceedings of the 35th IEEE/ACM International Conference on Automated Software Engineering},
pages = {90–102},
numpages = {13},
location = {Virtual Event, Australia},
series = {ASE '20}
}

@article{Tang2025documentation,
author = {Tang, Ye and Yan, Aoyang and Liu, Hui and Meng, Na and Zhong, Hao},
title = {Detecting Outdated Screenshot from GUI Document},
year = {2025},
publisher = {Association for Computing Machinery},
address = {New York, NY, USA},
issn = {1049-331X},
url = {https://doi.org/10.1145/3750041},
doi = {10.1145/3750041},
note = {Just Accepted},
journal = {ACM Trans. Softw. Eng. Methodol.},
month = jul
}

@article{Traini2021refactor,
author = {Traini, Luca and Di Pompeo, Daniele and Tucci, Michele and Lin, Bin and Scalabrino, Simone and Bavota, Gabriele and Lanza, Michele and Oliveto, Rocco and Cortellessa, Vittorio},
title = {How Software Refactoring Impacts Execution Time},
year = {2021},
issue_date = {April 2022},
publisher = {Association for Computing Machinery},
address = {New York, NY, USA},
volume = {31},
number = {2},
issn = {1049-331X},
url = {https://doi.org/10.1145/3485136},
doi = {10.1145/3485136},
journal = {ACM Trans. Softw. Eng. Methodol.},
month = dec,
articleno = {25},
numpages = {23}
}

@inproceedings{Lin2014Refactor,
author = {Lin, Yu and Radoi, Cosmin and Dig, Danny},
title = {Retrofitting concurrency for Android applications through refactoring},
year = {2014},
isbn = {9781450330565},
publisher = {Association for Computing Machinery},
address = {New York, NY, USA},
url = {https://doi.org/10.1145/2635868.2635903},
doi = {10.1145/2635868.2635903},
booktitle = {Proceedings of the 22nd ACM SIGSOFT International Symposium on Foundations of Software Engineering},
pages = {341–352},
numpages = {12},
location = {Hong Kong, China},
series = {FSE 2014}
}

@inproceedings{Khatchadourian2024Efficiency,
author = {Khatchadourian, Raffi and V\'{e}lez, Tatiana Castro and Bagherzadeh, Mehdi and Jia, Nan and Raja, Anita},
title = {Towards Safe Automated Refactoring of Imperative Deep Learning Programs to Graph Execution},
year = {2024},
isbn = {9798350329964},
publisher = {IEEE Press},
url = {https://doi.org/10.1109/ASE56229.2023.00187},
doi = {10.1109/ASE56229.2023.00187},
booktitle = {Proceedings of the 38th IEEE/ACM International Conference on Automated Software Engineering},
pages = {1800–1802},
numpages = {3},
location = {Echternach, Luxembourg},
series = {ASE '23}
}

@inproceedings{Qin2019Platform,
author = {Qin, Xue and Zhong, Hao and Wang, Xiaoyin},
title = {TestMig: migrating GUI test cases from iOS to Android},
year = {2019},
isbn = {9781450362245},
publisher = {Association for Computing Machinery},
address = {New York, NY, USA},
url = {https://doi.org/10.1145/3293882.3330575},
doi = {10.1145/3293882.3330575},
booktitle = {Proceedings of the 28th ACM SIGSOFT International Symposium on Software Testing and Analysis},
pages = {284–295},
numpages = {12},
location = {Beijing, China},
series = {ISSTA 2019}
}

@INPROCEEDINGS{Behrang2019GUI,
  author={Behrang, Farnaz and Orso, Alessandro},
  booktitle={2019 34th IEEE/ACM International Conference on Automated Software Engineering (ASE)}, 
  title={Test Migration Between Mobile Apps with Similar Functionality}, 
  year={2019},
  volume={},
  number={},
  pages={54-65},
  doi={10.1109/ASE.2019.00016}}

@INPROCEEDINGS{Talebipour2021GUI,
  author={Talebipour, Saghar and Zhao, Yixue and Dojcilović, Luka and Li, Chenggang and Medvidović, Nenad},
  booktitle={2021 36th IEEE/ACM International Conference on Automated Software Engineering (ASE)}, 
  title={UI Test Migration Across Mobile Platforms}, 
  year={2021},
  volume={},
  number={},
  pages={756-767},
  doi={10.1109/ASE51524.2021.9678643}}

@INPROCEEDINGS{Mukelabai2023Test,
  author={Mukelabai, Mukelabai and Derks, Christoph and Krüger, Jacob and Berger, Thorsten},
  booktitle={2023 38th IEEE/ACM International Conference on Automated Software Engineering (ASE)}, 
  title={To Share, or Not to Share: Exploring Test-Case Reusability in Fork Ecosystems}, 
  year={2023},
  volume={},
  number={},
  pages={837-849},
  doi={10.1109/ASE56229.2023.00191}}

@INPROCEEDINGS{Jiang2023Exploit,
  author={Jiang, Zheyue and Zhang, Yuan and Xu, Jun and Sun, Xinqian and Liu, Zhuang and Yang, Min},
  booktitle={2023 IEEE Symposium on Security and Privacy (SP)}, 
  title={AEM: Facilitating Cross-Version Exploitability Assessment of Linux Kernel Vulnerabilities}, 
  year={2023},
  volume={},
  number={},
  pages={2122-2137},
  doi={10.1109/SP46215.2023.10179286}}

@inproceedings{Chen2024Exploit,
author = {Chen, Zirui and Hu, Xing and Xia, Xin and Gao, Yi and Xu, Tongtong and Lo, David and Yang, Xiaohu},
title = {Exploiting Library Vulnerability via Migration Based Automating Test Generation},
year = {2024},
isbn = {9798400702174},
publisher = {Association for Computing Machinery},
address = {New York, NY, USA},
url = {https://doi.org/10.1145/3597503.3639583},
doi = {10.1145/3597503.3639583},
booktitle = {Proceedings of the IEEE/ACM 46th International Conference on Software Engineering},
articleno = {228},
numpages = {12},
location = {Lisbon, Portugal},
series = {ICSE '24}
}

@inproceedings{Zhou2024Exploit,
author = {Zhou, Zhuotong and Yang, Yongzhuo and Wu, Susheng and Huang, Yiheng and Chen, Bihuan and Peng, Xin},
title = {Magneto: A Step-Wise Approach to Exploit Vulnerabilities in Dependent Libraries via LLM-Empowered Directed Fuzzing},
year = {2024},
isbn = {9798400712487},
publisher = {Association for Computing Machinery},
address = {New York, NY, USA},
url = {https://doi.org/10.1145/3691620.3695531},
doi = {10.1145/3691620.3695531},
booktitle = {Proceedings of the 39th IEEE/ACM International Conference on Automated Software Engineering},
pages = {1633–1644},
numpages = {12},
location = {Sacramento, CA, USA},
series = {ASE '24}
}

@inproceedings{Bunyakiati2017Cherry,
author = {Bunyakiati, Panuchart and Phipathananunth, Chadarat},
title = {Cherry-picking of code commits in long-running, multi-release software},
year = {2017},
isbn = {9781450351058},
publisher = {Association for Computing Machinery},
address = {New York, NY, USA},
url = {https://doi.org/10.1145/3106237.3122818},
doi = {10.1145/3106237.3122818},
booktitle = {Proceedings of the 2017 11th Joint Meeting on Foundations of Software Engineering},
pages = {994–998},
numpages = {5},
location = {Paderborn, Germany},
series = {ESEC/FSE 2017}
}

@inproceedings{Ren2019Patch,
author = {Ren, Luyao},
title = {Automated patch porting across forked projects},
year = {2019},
isbn = {9781450355728},
publisher = {Association for Computing Machinery},
address = {New York, NY, USA},
url = {https://doi.org/10.1145/3338906.3342488},
doi = {10.1145/3338906.3342488},
booktitle = {Proceedings of the 2019 27th ACM Joint Meeting on European Software Engineering Conference and Symposium on the Foundations of Software Engineering},
pages = {1199–1201},
numpages = {3},
location = {Tallinn, Estonia},
series = {ESEC/FSE 2019}
}

@inproceedings{Pan2024Patch,
author = {Pan, Shengyi and Wang, You and Liu, Zhongxin and Hu, Xing and Xia, Xin and Li, Shanping},
title = {Automating Zero-Shot Patch Porting for Hard Forks},
year = {2024},
isbn = {9798400706127},
publisher = {Association for Computing Machinery},
address = {New York, NY, USA},
url = {https://doi.org/10.1145/3650212.3652134},
doi = {10.1145/3650212.3652134},
booktitle = {Proceedings of the 33rd ACM SIGSOFT International Symposium on Software Testing and Analysis},
pages = {363–375},
numpages = {13},
location = {Vienna, Austria},
series = {ISSTA 2024}
}

@inproceedings{Shariffdeen2021Patch,
author = {Shariffdeen, Ridwan and Gao, Xiang and Duck, Gregory J. and Tan, Shin Hwei and Lawall, Julia and Roychoudhury, Abhik},
title = {Automated patch backporting in Linux (experience paper)},
year = {2021},
isbn = {9781450384599},
publisher = {Association for Computing Machinery},
address = {New York, NY, USA},
url = {https://doi.org/10.1145/3460319.3464821},
doi = {10.1145/3460319.3464821},
booktitle = {Proceedings of the 30th ACM SIGSOFT International Symposium on Software Testing and Analysis},
pages = {633–645},
numpages = {13},
location = {Virtual, Denmark},
series = {ISSTA 2021}
}

@inproceedings{Yang2023Patch,
author = {Yang, Su and Xiao, Yang and Xu, Zhengzi and Sun, Chengyi and Ji, Chen and Zhang, Yuqing},
title = {Enhancing OSS Patch Backporting with Semantics},
year = {2023},
isbn = {9798400700507},
publisher = {Association for Computing Machinery},
address = {New York, NY, USA},
url = {https://doi.org/10.1145/3576915.3623188},
doi = {10.1145/3576915.3623188},
booktitle = {Proceedings of the 2023 ACM SIGSAC Conference on Computer and Communications Security},
pages = {2366–2380},
numpages = {15},
location = {Copenhagen, Denmark},
series = {CCS '23}
}

@inproceedings {Shi2023Patch,
author = {Youkun Shi and Yuan Zhang and Tianhan Luo and Xiangyu Mao and Yinzhi Cao and Ziwen Wang and Yudi Zhao and Zongan Huang and Min Yang},
title = {Backporting Security Patches of Web Applications: A Prototype Design and Implementation on Injection Vulnerability Patches},
booktitle = {31st USENIX Security Symposium (USENIX Security 22)},
year = {2022},
isbn = {978-1-939133-31-1},
address = {Boston, MA},
pages = {1993--2010},
url = {https://www.usenix.org/conference/usenixsecurity22/presentation/shi},
publisher = {USENIX Association},
month = aug
}

@inproceedings{Dai2021PoC,
author = {Dai, Jiarun and Zhang, Yuan and Xu, Hailong and Lyu, Haiming and Wu, Zicheng and Xing, Xinyu and Yang, Min},
title = {Facilitating Vulnerability Assessment through PoC Migration},
year = {2021},
isbn = {9781450384544},
publisher = {Association for Computing Machinery},
address = {New York, NY, USA},
url = {https://doi.org/10.1145/3460120.3484594},
doi = {10.1145/3460120.3484594},
booktitle = {Proceedings of the 2021 ACM SIGSAC Conference on Computer and Communications Security},
pages = {3300–3317},
numpages = {18},
location = {Virtual Event, Republic of Korea},
series = {CCS '21}
}

@INPROCEEDINGS{Lin2022GUI,
  author={Lin, Jun-Wei and Malek, Sam},
  booktitle={2022 IEEE Conference on Software Testing, Verification and Validation (ICST)}, 
  title={GUI Test Transfer from Web to Android}, 
  year={2022},
  volume={},
  number={},
  pages={1-11},
  doi={10.1109/ICST53961.2022.00011}}

@inproceedings{Gao2024Test,
author = {Gao, Yi and Hu, Xing and Xu, Tongtong and Xia, Xin and Lo, David and Yang, Xiaohu},
title = {MUT: Human-in-the-Loop Unit Test Migration},
year = {2024},
isbn = {9798400702174},
publisher = {Association for Computing Machinery},
address = {New York, NY, USA},
url = {https://doi.org/10.1145/3597503.3639124},
doi = {10.1145/3597503.3639124},
booktitle = {Proceedings of the IEEE/ACM 46th International Conference on Software Engineering},
articleno = {229},
numpages = {12},
location = {Lisbon, Portugal},
series = {ICSE '24}
}

@ARTICLE{Sondhi2022Test,
  author={Sondhi, Devika and Jobanputra, Mayank and Rani, Divya and Purandare, Salil and Sharma, Sakshi and Purandare, Rahul},
  journal={IEEE Transactions on Software Engineering}, 
  title={Mining Similar Methods for Test Adaptation}, 
  year={2022},
  volume={48},
  number={7},
  pages={2262-2276},
  doi={10.1109/TSE.2021.3057163}}

@inproceedings{Zhang2024GUI,
author = {Zhang, Yakun and Zhu, Qihao and Yan, Jiwei and Liu, Chen and Zhang, Wenjie and Zhao, Yifan and Hao, Dan and Zhang, Lu},
title = {Synthesis-Based Enhancement for GUI Test Case Migration},
year = {2024},
isbn = {9798400706127},
publisher = {Association for Computing Machinery},
address = {New York, NY, USA},
url = {https://doi.org/10.1145/3650212.3680327},
doi = {10.1145/3650212.3680327},
booktitle = {Proceedings of the 33rd ACM SIGSOFT International Symposium on Software Testing and Analysis},
pages = {869–881},
numpages = {13},
location = {Vienna, Austria},
series = {ISSTA 2024}
}

@inproceedings{Zhang2024GUI2,
author = {Zhang, Yakun and Zhang, Wenjie and Ran, Dezhi and Zhu, Qihao and Dou, Chengfeng and Hao, Dan and Xie, Tao and Zhang, Lu},
title = {Learning-based Widget Matching for Migrating GUI Test Cases},
year = {2024},
isbn = {9798400702174},
publisher = {Association for Computing Machinery},
address = {New York, NY, USA},
url = {https://doi.org/10.1145/3597503.3623322},
doi = {10.1145/3597503.3623322},
booktitle = {Proceedings of the IEEE/ACM 46th International Conference on Software Engineering},
articleno = {69},
numpages = {13},
location = {Lisbon, Portugal},
series = {ICSE '24}
}

@INPROCEEDINGS{Nielsen2021Patch,
  author={Nielsen, Benjamin Barslev and Torp, Martin Toldam and Møller, Anders},
  booktitle={2021 IEEE/ACM 43rd International Conference on Software Engineering (ICSE)}, 
  title={Semantic Patches for Adaptation of JavaScript Programs to Evolving Libraries}, 
  year={2021},
  volume={},
  number={},
  pages={74-85},
  doi={10.1109/ICSE43902.2021.00020}}

@INPROCEEDINGS{Mujahid2023Library,
  author={Mujahid, Suhaib and Costa, Diego Elias and Abdalkareem, Rabe and Shihab, Emad},
  booktitle={2023 38th IEEE/ACM International Conference on Automated Software Engineering (ASE)}, 
  title={Where to Go Now? Finding Alternatives for Declining Packages in the npm Ecosystem}, 
  year={2023},
  volume={},
  number={},
  pages={1628-1639},
  doi={10.1109/ASE56229.2023.00119}}

@INPROCEEDINGS{Alrubaye2019Library,
  author={Alrubaye, Hussein and Mkaouer, Mohamed Wiem and Ouni, Ali},
  booktitle={2019 IEEE/ACM 27th International Conference on Program Comprehension (ICPC)}, 
  title={On the Use of Information Retrieval to Automate the Detection of Third-Party Java Library Migration at the Method Level}, 
  year={2019},
  volume={},
  number={},
  pages={347-357},
  doi={10.1109/ICPC.2019.00053}}

@article{Huang2024API,
author = {Huang, Zhenfei and Chen, Junjie and Jiang, Jiajun and Liang, Yihua and You, Hanmo and Li, Fengjie},
title = {Mapping APIs in Dynamic-typed Programs by Leveraging Transfer Learning},
year = {2024},
issue_date = {May 2024},
publisher = {Association for Computing Machinery},
address = {New York, NY, USA},
volume = {33},
number = {4},
issn = {1049-331X},
url = {https://doi.org/10.1145/3641848},
doi = {10.1145/3641848},
journal = {ACM Trans. Softw. Eng. Methodol.},
month = apr,
articleno = {102},
numpages = {29}
}

@INPROCEEDINGS{Zhu2024API,
  author={Zhu, Chenguang and Saha, Ripon K. and Prasad, Mukul R. and Khurshid, Sarfraz},
  booktitle={2021 36th IEEE/ACM International Conference on Automated Software Engineering (ASE)}, 
  title={Restoring the Executability of Jupyter Notebooks by Automatic Upgrade of Deprecated APIs}, 
  year={2021},
  volume={},
  number={},
  pages={240-252},
  doi={10.1109/ASE51524.2021.9678889}}

@inproceedings{Ramos2023API,
author = {Ramos, Daniel and Mitchell, Hailie and Lynce, In\^{e}s and Manquinho, Vasco and Martins, Ruben and Goues, Claire Le},
title = {MELT: Mining Effective Lightweight Transformations from Pull Requests},
year = {2024},
isbn = {9798350329964},
publisher = {IEEE Press},
url = {https://doi.org/10.1109/ASE56229.2023.00117},
doi = {10.1109/ASE56229.2023.00117},
booktitle = {Proceedings of the 38th IEEE/ACM International Conference on Automated Software Engineering},
pages = {1516–1528},
numpages = {13},
location = {Echternach, Luxembourg},
series = {ASE '23}
}

@ARTICLE{Lamothe2022API ,
  author={Lamothe, Maxime and Shang, Weiyi and Chen, Tse-Hsun Peter},
  journal={IEEE Transactions on Software Engineering}, 
  title={A3: Assisting Android API Migrations Using Code Examples}, 
  year={2022},
  volume={48},
  number={2},
  pages={417-431},
  doi={10.1109/TSE.2020.2988396}}

@inproceedings{zou2024syzbridge,
  title={SyzBridge: Bridging the Gap in Exploitability Assessment of Linux Kernel Bugs in the Linux Ecosystem},
  author={Zou, Xiaochen and Hao, Yu and Zhang, Zheng and Pu, Juefei and Chen, Weiteng and Qian, Zhiyun},
  booktitle={31st Annual Network and Distributed System Security Symposium, NDSS},
  year={2024}
}

@inproceedings{Fazzini2019API,
author = {Fazzini, Mattia and Xin, Qi and Orso, Alessandro},
title = {Automated API-usage update for Android apps},
year = {2019},
isbn = {9781450362245},
publisher = {Association for Computing Machinery},
address = {New York, NY, USA},
url = {https://doi.org/10.1145/3293882.3330571},
doi = {10.1145/3293882.3330571},
booktitle = {Proceedings of the 28th ACM SIGSOFT International Symposium on Software Testing and Analysis},
pages = {204–215},
numpages = {12},
location = {Beijing, China},
series = {ISSTA 2019}
}

@INPROCEEDINGS{Nguyen2019API,
  author={Nguyen, Phuong T. and Di Rocco, Juri and Di Ruscio, Davide and Ochoa, Lina and Degueule, Thomas and Di Penta, Massimiliano},
  booktitle={2019 IEEE/ACM 41st International Conference on Software Engineering (ICSE)}, 
  title={FOCUS: A Recommender System for Mining API Function Calls and Usage Patterns}, 
  year={2019},
  volume={},
  number={},
  pages={1050-1060},
  doi={10.1109/ICSE.2019.00109}}

@INPROCEEDINGS{Xu2019API,
  author={Xu, Shengzhe and Dong, Ziqi and Meng, Na},
  booktitle={2019 IEEE/ACM 27th International Conference on Program Comprehension (ICPC)}, 
  title={Meditor: Inference and Application of API Migration Edits}, 
  year={2019},
  volume={},
  number={},
  pages={335-346},
  doi={10.1109/ICPC.2019.00052}}

@inproceedings{Linares2013Time,
author = {Linares-V\'{a}squez, Mario and Bavota, Gabriele and Bernal-C\'{a}rdenas, Carlos and Di Penta, Massimiliano and Oliveto, Rocco and Poshyvanyk, Denys},
title = {API change and fault proneness: a threat to the success of Android apps},
year = {2013},
isbn = {9781450322379},
publisher = {Association for Computing Machinery},
address = {New York, NY, USA},
url = {https://doi.org/10.1145/2491411.2491428},
doi = {10.1145/2491411.2491428},
booktitle = {Proceedings of the 2013 9th Joint Meeting on Foundations of Software Engineering},
pages = {477–487},
numpages = {11},
location = {Saint Petersburg, Russia},
series = {ESEC/FSE 2013}
}

@INPROCEEDINGS{Lamothe2018Time,
  author={Lamothe, Maxime and Shang, Weiyi},
  booktitle={2018 IEEE/ACM 15th International Conference on Mining Software Repositories (MSR)}, 
  title={Exploring the Use of Automated API Migrating Techniques in Practice: An Experience Report on Android}, 
  year={2018},
  volume={},
  number={},
  pages={503-514},
  doi={}}

@inproceedings{Bennett2000Maintenance,
author = {Bennett, Keith H. and Rajlich, V\'{a}clav T.},
title = {Software maintenance and evolution: a roadmap},
year = {2000},
isbn = {1581132530},
publisher = {Association for Computing Machinery},
address = {New York, NY, USA},
url = {https://doi.org/10.1145/336512.336534},
doi = {10.1145/336512.336534},
booktitle = {Proceedings of the Conference on The Future of Software Engineering},
pages = {73–87},
numpages = {15},
location = {Limerick, Ireland},
series = {ICSE '00}
}

@ARTICLE{Rajlich2000life,
  author={Rajlich, V.T. and Bennett, K.H.},
  journal={Computer}, 
  title={A staged model for the software life cycle}, 
  year={2000},
  volume={33},
  number={7},
  pages={66-71},
  doi={10.1109/2.869374}}

@inproceedings{Fazzini2020Address,
author = {Fazzini, Mattia and Xin, Qi and Orso, Alessandro},
title = {APIMigrator: an API-usage migration tool for Android apps},
year = {2020},
isbn = {9781450379595},
publisher = {Association for Computing Machinery},
address = {New York, NY, USA},
url = {https://doi.org/10.1145/3387905.3388608},
doi = {10.1145/3387905.3388608},
booktitle = {Proceedings of the IEEE/ACM 7th International Conference on Mobile Software Engineering and Systems},
pages = {77–80},
numpages = {4},
location = {Seoul, Republic of Korea},
series = {MOBILESoft '20}
}

@inproceedings{Vanover2020Test,
author = {Vanover, Jackson and Deng, Xuan and Rubio-Gonz\'{a}lez, Cindy},
title = {Discovering discrepancies in numerical libraries},
year = {2020},
isbn = {9781450380089},
publisher = {Association for Computing Machinery},
address = {New York, NY, USA},
url = {https://doi.org/10.1145/3395363.3397380},
doi = {10.1145/3395363.3397380},
booktitle = {Proceedings of the 29th ACM SIGSOFT International Symposium on Software Testing and Analysis},
pages = {488–501},
numpages = {14},
location = {Virtual Event, USA},
series = {ISSTA 2020}
}
